\documentclass[nofootinbib,10pt,twocolumn,pra,showpacs,tightenlines, showkeys, floatfix, english,aps]{revtex4}
\usepackage[T1]{fontenc}
\usepackage[latin9]{inputenc}
\usepackage{color}
\usepackage{float}
\usepackage{epsfig,graphicx,times}
\usepackage{amstext}
\usepackage{amsmath}  
\usepackage{amssymb}
\usepackage{stackrel}
\usepackage{graphicx}
\usepackage{esint}
\usepackage{epstopdf}
\usepackage{epsfig}
\usepackage{bm}

\makeatother

\usepackage{babel}
\begin{document}

\title{Linear and nonlinear quantum Zeno and anti-Zeno effects in a nonlinear optical coupler}

\author{Kishore Thapliyal$^{a}$, Anirban Pathak$^{a,}$\footnote{Email: anirban.pathak@gmail.com, Phone: +91 9717066494}, and Jan ${\rm \check{Perina}}$$^{b,c}$}

\affiliation{$^{a}$Jaypee Institute of Information Technology, A-10, Sector-62,
Noida, UP-201307, India\\
$^{b}$RCPTM, Joint Laboratory of Optics of Palacky University and
Institute of Physics of Academy of Science of the Czech Republic,
Faculty of Science, Palacky University, 17. listopadu 12, 771 46 Olomouc,
Czech Republic\\
$^{c}$Department of Optics, Palacky University, 17. listopadu 12,
771 46 Olomouc, Czech Republic}
\begin{abstract}
Quantum Zeno and anti-Zeno effects are studied in a symmetric nonlinear
optical coupler, which is composed of two nonlinear ($\chi^{\left(2\right)}$)
waveguides that are interacting with each other via the evanescent
waves. Both the waveguides operate under second harmonic generation.
However, to study quantum Zeno and anti-Zeno effects one of them is
considered as the system and the other one is considered as the probe.
Considering all the fields involved as weak, a completely quantum
mechanical description is provided, and the analytic solutions of
Heisenberg's equations of motion for all the field modes are obtained
using a perturbative technique.  Photon number statistics of the second
harmonic mode of the system is shown to depend on the presence of
the probe, and this dependence is considered as quantum Zeno and anti-Zeno
effects. Further, it is established that as a special case of the
momentum operator for $\chi^{\left(2\right)}-\chi^{\left(2\right)}$
symmetric coupler we can obtain momentum operator of $\chi^{\left(2\right)}-\chi^{\left(1\right)}$
asymmetric coupler with linear ($\chi^{\left(1\right)}$) waveguide
as the probe, and in such a particular case, the expressions obtained
for Zeno and anti-Zeno effects with nonlinear probe (which we referred
to as nonlinear quantum Zeno and anti-Zeno effects) may be reduced
to the corresponding expressions with linear probe (which we referred
to as the linear quantum Zeno and anti-Zeno effects). Linear and nonlinear
quantum Zeno and anti-Zeno effects are rigorously investigated, and
it is established that in the stimulated case, we may switch between
quantum Zeno and anti-Zeno effects just by controlling the phase of
the second harmonic mode of the system or probe. 
\end{abstract}

\maketitle

\section{Introduction\label{sec:Introduction}}

In the 5th century BC, the Greek philosopher
Zeno of Elea introduced a set of paradoxes of motion. These paradoxes,
which are now known as Zeno's paradoxes, were unsolved for long, and
they fascinated mathematicians, logicians, physicists and other creative
minds since their introduction. In the recent past, a quantum analogue
of Zeno's paradoxes has been studied intensively. Specifically, in
the late 50s and early 60s of the 20th century, Khalfin studied nonexponential
decay of unstable atoms \cite{Khalfin}. Later on, in 1977, Misra
and Sudarshan \cite{misra1977zeno} showed that under continuous measurement,
an unstable particle will never be found to decay, and in analogy
with classical Zeno's paradox they named this phenomenon as \emph{Zeno's
quantum paradox}. Quantum Zeno effect (QZE) in the original formulation
refers to the inhibition of the temporal evolution of a system on
continuous measurement \cite{misra1977zeno}, while quantum anti-Zeno
effect (QAZE) or inverse Zeno effect refers the enhancement of the
evolution instead of the inhibition (see Refs. \cite{venugopalan2007zeno_review,facchi2001zeno-review,saverio-zeno in 70 minutes}
for the reviews). Here, it is important to note that usually quantum
Zeno effect is viewed as a process, which is associated with the repeated
projective measurement. This is only a specific manifestation of quantum
Zeno effect. In fact, it can be manifested in a few equivalent ways
\cite{saverio 3 manifestation}. One such manifestation of quantum
Zeno effect is a process in which continuous interaction between the
system and probe leads to quantum Zeno effect. Here, we aim to study
the continuous interaction type manifestation of quantum Zeno effect
in a symmetric nonlinear optical coupler, which is made of two nonlinear
waveguides with $\chi^{(2)}$ nonlinearity, and each of the waveguides
is operating under second harmonic generation. As we are describing
an optical coupler, the waveguides are coupled with each other. More
precisely, these two waveguides interact with each other through the
evanescent waves. We consider one of the waveguides as the probe and
the other one as the system. In what follows, we will show that the
beauty of the symmetric nonlinear optical coupler ($\chi^{(2)}-\chi^{(2)}$
coupler) studied here is that the results obtained for this coupler
can be directly reduced to the corresponding results for an asymmetric
nonlinear optical coupler where the probe is linear ($\chi^{(1)})$
and the system is nonlinear ($\chi^{(2)}).$ We have reported quantum
Zeno and anti-Zeno effects in both the symmetric ($\chi^{(2)}-\chi^{(2)}$)
nonlinear optical coupler and asymmetric ($\chi^{(2)}-\chi^{(1)})$
nonlinear optical coupler. Following an earlier work \cite{nonlinear-zeno-name-cpoupler-Abdullaev},
we refer to the Zeno effect observed due to the continuous interaction
of a nonlinear $\left(\chi^{(2)}\right)$ probe as the nonlinear Zeno
effect. Similarly, the Zeno effect observed due to the continuous
interaction of a linear $\left(\chi^{(1)}\right)$ probe is referred
to as the linear Zeno effect.

Optical couplers can be prepared easily and several exciting applications
of the optical couplers have been reported in the recent past (see
\cite{kishore2014co-coupler,kishore2014contra,perina-rev} and references therein).
Consequently, it is no wonder that quantum Zeno and anti-Zeno effects
have been investigated in various types of optical couplers \cite{Rechacek-2001-zrno-coupler,thun2002zeno-raman,chi2-chi1-spie,chi2-chi2,rehacek2000zeno-coupler}. Specifically, quantum Zeno and anti-Zeno effects were shown in Raman
and Brillouin scattering using a ($\chi^{(3)}-\chi^{(1)}$) asymmetric
nonlinear optical coupler \cite{thun2002zeno-raman}; their existence
was also shown in the $\chi^{(2)}-\chi^{(1)}$ optical couplers \cite{chi2-chi1-spie},
$\chi^{(2)}-\chi^{(2)}$ optical couplers \cite{chi2-chi2}, etc.
In all these studies, it was always considered that one of the mode
in the nonlinear waveguide (this waveguide is considered as the system)
is coupled with the auxiliary mode in a (non)linear waveguide (this
waveguide is considered as the probe). Actually, the auxiliary mode
acts as the probe since its coupling with the system implements continuous
observation on the evolution of the system (nonlinear waveguide) and
changes the photon statistics of the other modes (which are not coupled
to the probe mode) of the nonlinear waveguide. Quantum Zeno and anti-Zeno
effects have also been investigated in optical systems other than
couplers, such as in parametric down-conversion \cite{down-conversion-perina2,parametric-down-conversion-anti-zeno,down-conversion-perina},
parametric down conversion with losses \cite{perina-zeno-parametric-dc},
an arrangement of beam splitters \cite{All-optical-zeno-agarwal},
etc. In these studies on quantum Zeno effect in optical systems, often
the pump mode has been considered strong, and thus the complexity
of a completely quantum mechanical treatment has been circumvented.
Keeping this in mind, here we plan to use a completely quantum mechanical
description of the coupler.

Initially, interest in quantum Zeno effect was theoretical and purely
academic in nature, but with time quantum Zeno effect has been experimentally
realized by several groups using different techniques \cite{kwait-int.-free.measurement-expt.,experimental-zeno-1,experimental-zeno-2}.
Not only that, several interesting applications of quantum Zeno effect
have also been proposed \cite{kwait-int.-free.measurement-expt.,counterfactual-quantum computation,Zubairy1,zeno-tomography-hradil,zeno-tomography2}.
Specifically, in Refs. \cite{zeno-tomography-hradil,zeno-tomography2},
it was established that the quantum Zeno effect may be used to increase
the resolution of absorption tomography. A few of the proposed applications
have also been experimentally realized. For example, Kwait et al.
implemented high-efficiency quantum interrogation measurement using
quantum Zeno effect \cite{kwait-int.-free.measurement-expt.}. Until
recently, all the investigations related to the quantum Zeno effect
were restricted to the microscopic world. Recently, in a very interesting
work, it has been extended to the macroscopic world by showing the
evidence for the existence of quantum Zeno effect for large black
holes \cite{Zeno-black hole}. Possibility of observing the macroscopic
Zeno effect was also studied in the context of stationary flows in
nonlinear waveguides with localized dissipation \cite{macroscopic zeno PRL}.
The interest in quantum Zeno effect has recently been amplified with
the advent of various protocols of quantum communication that are
based on quantum Zeno effect. Specifically, in Ref. \cite{Zubairy1},
a counterfactual protocol of direct quantum communication was proposed
using chained quantum Zeno effect, and in Ref. \cite{counterfactual-quantum computation},
the same effect is used to propose a scheme for counterfactual quantum
computation. In the past, a proposal for quantum computing was made
using an environment induced quantum Zeno effect to confine the dynamics
in a decoherence-free subspace \cite{Decoherence-free}. Recently,
quantum Zeno effect has also been used to reduce communication complexity
\cite{Q-com-comp}. These applications of quantum Zeno effect and
easy production of optical couplers motivated us to systematically
investigate the possibility of observing quantum Zeno and anti-Zeno
effects in a symmetric nonlinear coupler which is not studied earlier
using a completely quantum description.

To investigate the existence of quantum Zeno and anti-Zeno effects
in the optical coupler of our interest, we have obtained closed form
analytic expressions for the spatial evolution of the different field
operators using the Sen-Mandal perturbative approach \cite{kishore2014co-coupler,kishore2014contra,mandal2004co-coupler},
which is known to produce better results compared to the usual short-length
approximation method \cite{perina1995photon}. Actually, in sharp
contrast to the short length approximated solutions, the solutions
of Heisenberg's equations of motion obtained using the Sen-Mandal
approach are not restricted by length. This is why we use Sen-Mandal
perturbative approach and a completely quantum mechanical description
of the coupler for our investigation. In the past, photon statistics
and dynamics of the symmetric coupler of our interest was studied
by some of the present authors with an assumption that both the second
harmonic modes are strong \cite{sym-coup}. The assumption circumvented
the use of completely quantum mechanical description. Further, the
system has also been used to model a beam splitter with second order
nonlinearity \cite{coup-bs}. Present investigation, not only revealed
the existence of nonlinear quantum Zeno and anti-Zeno effects it also
established the existence of linear Zeno and anti-Zeno effects. The
study also showed that switching between quantum Zeno and anti-Zeno
effects is possible by varying phase-mismatches.

The rest of the paper is organized as follows. In Section \ref{sec:System-and-solution},
we briefly describe the momentum operator for the symmetric nonlinear
optical coupler and the method used here to obtain the analytic expressions
of the spatial evolution of the field operators of various modes.
However, the detailed solution obtained here is shown in the Appendix
A. In Section \ref{sec:Linear-and-nonlinear}, the existence of quantum
Zeno and anti-Zeno effects are systematically investigated. Finally
the paper is concluded in Section \ref{sec:Conclusions}.

\section{System and solution\label{sec:System-and-solution}}

\begin{widetext}

\begin{figure}[h]
\includegraphics[angle=-90,scale=0.5]{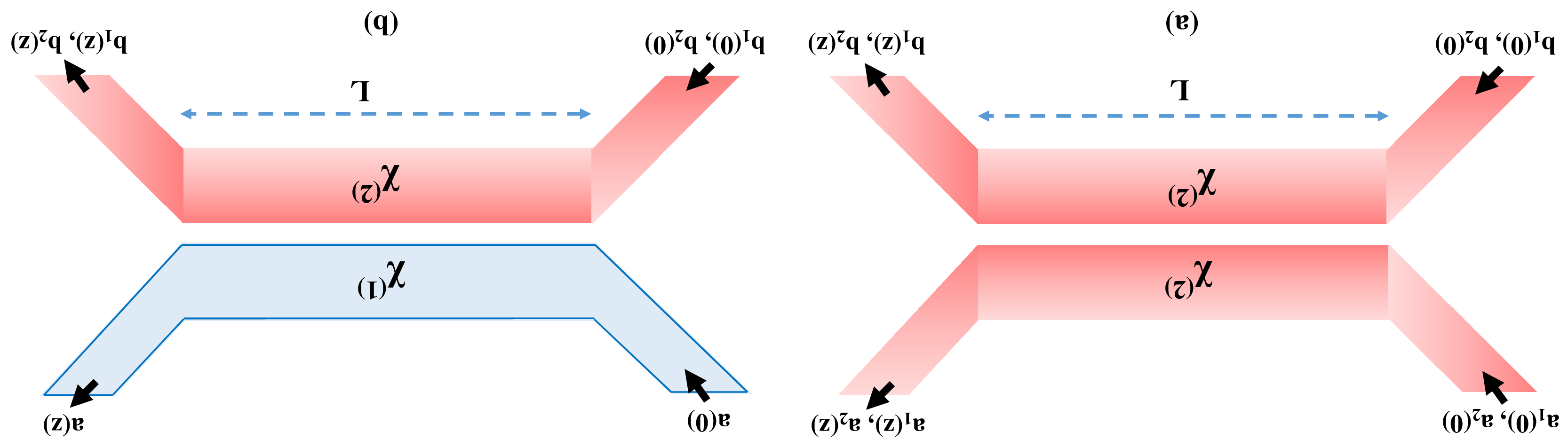}
\protect\caption{\label{fig:Schematic-diagram}(Color online) Schematic diagrams of
(a) a symmetric and (b) an asymmetric nonlinear optical coupler of
interaction length $L$ in a codirectional propagation of different
field modes involved. The symmetric coupler is prepared by combining
two nonlinear (quadratic) waveguides operating under second harmonic
generation, and in the asymmetric coupler one nonlinear waveguide
of the symmetric coupler is replaced by a linear waveguide.}
\end{figure}

\end{widetext}

Momentum operator of a symmetric nonlinear optical coupler, prepared
by combining two nonlinear (quadratic) waveguides operated by second
harmonic generation (as shown in Fig. \ref{fig:Schematic-diagram}
a), in interaction picture is given by \cite{hamiltonian}
\begin{equation}
\begin{array}{lcl}
G_{{\rm sym}} & = & \hbar ka_{1}b_{1}^{\dagger}+\hbar\Gamma_{a}a_{1}^{2}a_{2}^{\dagger}\exp(i\Delta k_{a}z)\\
 & + & \hbar\Gamma_{b}b_{1}^{2}b_{2}^{\dagger}\exp(i\Delta k_{b}z)\,+{\rm H.c}.\,,
\end{array}\label{eq:Ham-symmetric}
\end{equation}
 where the annihilation (creation) operators $a_{i}\,(a_{i}^{\dagger})$
and $b_{i}\,(b_{i}^{\dagger})$ correspond to the field operators
in two nonlinear waveguides. Here, $a_{1}(k_{a_{1}})$ and $a_{2}(k_{a_{2}})$
denote annihilation operators (wave vectors) for fundamental and second
harmonic modes, respectively, in one waveguide. Similarly, $b_{1}(k_{b_{1}})$
and $b_{2}(k_{b_{2}})$ represent annihilation operators (wave vectors)
for fundamental and second harmonic modes, respectively, in second
waveguide. Further, ${\rm H.c.}$ stands for the Hermitian conjugate;
$\Delta k_{j}=|2k_{j_{1}}-k_{j_{2}}|$ refers to the phase mismatch
between the fundamental and second harmonic beams; the parameters
$k$ and $\Gamma_{j}$ denote the linear and nonlinear coupling constants,
respectively, where $j\in\left\{ a,b\right\} $. The momentum operator
described above is completely quantum mechanical in the sense that
all the modes involved in the process are considered weak and treated
quantum mechanically. Thus, we consider pump as weak and note that
$\Gamma_{j}\ll k$ as $k$ and $\Gamma_{j}$ are proportional to the
linear $(\chi^{(1)})$ and nonlinear $(\chi^{(2)})$ susceptibilities,
respectively and usually $\chi^{(2)}/\chi^{(1)}\,\simeq10^{-6}$.

For the study of quantum Zeno and anti-Zeno effects in a system we
need a system momentum operator with a continuous interaction with
a probe. In this particular system, the symmetric nonlinear optical
coupler, we can consider that the system, which is described by $G_{{\rm sys}}=\hbar\Gamma_{b}b_{1}^{2}b_{2}^{\dagger}\exp(i\Delta k_{b}z)\,+{\rm H.c}.\,$,
is in continuous interaction with the probe, which is described by
$G_{{\rm probe}}=\hbar ka_{1}b_{1}^{\dagger}+\hbar\Gamma_{a}a_{1}^{2}a_{2}^{\dagger}\exp(i\Delta k_{a}z)\,+{\rm H.c}..$
Here, the probe itself is considered to be nonlinear. Further, if
we take $\Gamma_{a}=0$ in Eq. (\ref{eq:Ham-symmetric}), i.e., if
we consider probe to be linear, we obtain \cite{kishore2014co-coupler,kishore2014contra,mandal2004co-coupler}
\begin{equation}
G_{{\rm asym}}=\hbar kab_{1}^{\dagger}+\hbar\Gamma b_{1}^{2}b_{2}^{\dagger}\exp(i\Delta kz)\,+{\rm H.c}.\,,\label{eq:Ham-asymmetric}
\end{equation}
which is the momentum operator of an asymmetric nonlinear optical
coupler in the interaction picture. 

The spatial evolution of various modes involved in the momentum operators
(\ref{eq:Ham-symmetric}-\ref{eq:Ham-asymmetric}) can be obtained
as the simultaneous solutions of the Heisenberg's equations of motion
corresponding to each mode. However, for the complex systems, such
as considered here, the closed form analytic solutions are possible
only by using some perturbative methods. Here, we use Sen-Mandal perturbative
method \cite{kishore2014co-coupler,kishore2014contra,chi2-chi1-spie,mandal2004co-coupler},
which has already been shown to be superior to the frequently used
short-length/time method \cite{thun2002zeno-raman,perina1995photon}.
In fact, the solutions obtained using the short-length perturbative
technique can be obtained as a limiting case of a solution obtained
using Sen-Mandal approach. Specifically, it may be obtained by neglecting
higher power terms in length, from the solutions of the Sen-Mandal
method. Actually, in Sen-Mandal approach, potential solutions of the
Heisenberg equations of motion for different field modes are systematically
constructed. The assumed solution for the evolution of annihilation
operator of a field mode is constructed in such a way that it contains
all the possible higher power terms in length, but higher power terms
in weak coupling constant are neglected (for detail see \cite{kishore2014co-coupler,kishore2014contra,chi2-chi1-spie,mandal2004co-coupler}).
As the solutions obtained using Sen-Mandal method is applicable to
relatively large interaction lengths and it contains several higher
power terms that are neglected in conventional short-length approach,
it provides more accurate solution compared to the short-length solutions.
Further, the solution obtained using Sen-Mandal method is often found
successful in detecting nonclassical character of a physical system
that are not detected by conventional short- length/time solution \cite{thun2002zeno-raman,perina1995photon}. Keeping these facts
in mind, here we use Sen-Mandal method to obtain spatial evolution
of all the field operators involved in (\ref{eq:Ham-symmetric}).
However, the closed form analytic
expresseion for the spatial evolution in $b_{2}$ mode, i.e., $b_{2}(z)$
is provided in Appendix A. In the Appendix A, we restrict our description
to $b_{2}(z)$ as only the coefficeints appear in the analytic expression
of $b_{2}(z)$ appear in the expressions of linear and nonlinear Zeno
parameters. Specifically, Heisenberg's equations of motion for different
field modes involved in the momentum operator (\ref{eq:Ham-symmetric})
are obtained in Eq. (\ref{eq:Heisenberg's-eqs}) in Appendix A. The closed form analytic expressions
for the spatial evolution of all the field modes up to quadratic terms
in nonlinear coupling constants ($\Gamma_{j}$) are subsequently obtained
using Sen-Mandal perturbation method (cf. Suuplementary material for
all modes and Appendix A for $b_{2}$ modes). In
what follows, we use the expression of $b_{2}(z)$ provided in Appendix
A to investigate the linear and nonlinear Zeno and anti-Zeno effects
in the optical couplers.

\section{Linear and nonlinear quantum Zeno and anti-Zeno effects \label{sec:Linear-and-nonlinear}}

Being consistent with the theme of the present work, the presence
of quantum Zeno and anti-Zeno effects with a nonlinear probe corresponds
to the nonlinear quantum Zeno and anti-Zeno effects. Similarly, a
linear probe will give the linear quantum Zeno and anti-Zeno effects.
Further, it has already been mentioned in Section (\ref{eq:Ham-symmetric}),
that analytical expressions for Zeno parameter for a linear probe
can be obtained as the limiting cases of the expressions obtained
for nonlinear probe by neglecting the nonlinearity present in the
probe \cite{non-QZeno}. Quite similar analogue of nonlinear and linear
quantum Zeno and anti-Zeno effects were also discussed in the recent
past \cite{nonlinear-zeno-name-cpoupler-Abdullaev,non-Zeno} in other
physical systems.

\subsection{Number operator and Zeno parameter}

The analytic expression of the number operator for the second harmonic
field mode in the system waveguide, i.e., $b_{2}$ mode using Eq.
(\ref{eq:assumed-sol}) in Appendix A is given by
\begin{widetext}
\begin{equation}
\begin{array}{lcl}
N_{b_{2}}\left(z\right) & = & b_{2}^{\dagger}\left(z\right)b_{2}\left(z\right)\\
 & = & b_{2}^{\dagger}\left(0\right)b_{2}\left(0\right)+\left|l_{2}\right|^{2}b_{1}^{\dagger2}\left(0\right)b_{1}^{2}\left(0\right)+\left|l_{3}\right|^{2}a_{1}^{\dagger}\left(0\right)a_{1}\left(0\right)b_{1}^{\dagger}\left(0\right)b_{1}\left(0\right)+\left|l_{4}\right|^{2}a_{1}^{\dagger2}\left(0\right)a_{1}^{2}\left(0\right)+\left[l_{2}b_{2}^{\dagger}\left(0\right)b_{1}^{2}\left(0\right)\right.\\
 & + & l_{3}b_{2}^{\dagger}\left(0\right)b_{1}\left(0\right)a_{1}\left(0\right)+l_{4}b_{2}^{\dagger}\left(0\right)a_{1}^{2}\left(0\right)+l_{2}^{*}l_{3}a_{1}\left(0\right)b_{1}^{\dagger2}\left(0\right)b_{1}\left(0\right)+l_{2}^{*}l_{4}a_{1}^{2}\left(0\right)b_{1}^{\dagger2}\left(0\right)+l_{3}^{*}l_{4}a_{1}^{\dagger}\left(0\right)a_{1}^{2}\left(0\right)b_{1}^{\dagger}\left(0\right)\\
 & + & l_{5}b_{1}^{\dagger}\left(0\right)b_{1}\left(0\right)b_{2}^{\dagger}\left(0\right)b_{2}\left(0\right)+l_{6}b_{2}^{\dagger}\left(0\right)b_{2}\left(0\right)+l_{7}a_{1}\left(0\right)b_{1}^{\dagger}\left(0\right)b_{2}^{\dagger}\left(0\right)b_{2}\left(0\right)+l_{8}a_{1}^{\dagger}\left(0\right)b_{1}\left(0\right)b_{2}^{\dagger}\left(0\right)b_{2}\left(0\right)\\
 & + & l_{9}a_{1}^{\dagger}\left(0\right)a_{1}\left(0\right)b_{2}^{\dagger}\left(0\right)b_{2}\left(0\right)+l_{10}a_{1}^{\dagger}\left(0\right)a_{2}\left(0\right)b_{1}\left(0\right)b_{2}^{\dagger}\left(0\right)+l_{11}a_{1}^{\dagger}\left(0\right)a_{1}\left(0\right)a_{2}\left(0\right)b_{2}^{\dagger}\left(0\right)+l_{12}a_{2}\left(0\right)b_{2}^{\dagger}\left(0\right)\\
 & + & \left.l_{13}a_{2}\left(0\right)b_{1}^{\dagger}\left(0\right)b_{1}\left(0\right)b_{2}^{\dagger}\left(0\right)+l_{14}a_{1}\left(0\right)a_{2}\left(0\right)b_{1}^{\dagger}\left(0\right)b_{2}^{\dagger}\left(0\right)+{\rm H.c.}\right],
\end{array}\label{eq:nb2}
\end{equation}
\end{widetext}
where the functional form of coefficients $l_{i}$ is given in Eq.
(\ref{eq:coefficient-l}) in Appendix A. 

Without any loss of generality, we considered the initial state being
a multimode coherent state $|\alpha\rangle|\beta\rangle|\gamma\rangle|\delta\rangle$,
which is the product of four single mode coherent states $|\alpha\rangle,\,|\beta\rangle$,
$|\gamma\rangle$, and $|\delta\rangle$. Here, $|\alpha\rangle,\,|\beta\rangle$,
$|\gamma\rangle$, and $|\delta\rangle$ are the eigenkets of the
annihilation operators for the corresponding field modes, i.e., $a_{1}$,
$b_{1}$, $a_{2}$, and $b_{2}$, respectively. For example, $b_{1}(0)|\alpha\rangle|\beta\rangle|\gamma\rangle|\delta\rangle=\beta|\alpha\rangle|\beta\rangle|\gamma\rangle|\delta\rangle$
and $a_{1}(0)|\alpha\rangle|\beta\rangle|\gamma\rangle|\delta\rangle=\alpha|\alpha\rangle|\beta\rangle|\gamma\rangle|\delta\rangle,$
where $|\alpha|^{2},\,|\beta|^{2},\,|\gamma|^{2},$ and $|\delta|^{2}$
are the initial number of photons in the field modes $a_{1}$, $b_{1}$,
$a_{2}$, and $b_{2}$, respectively. The symmetric nonlinear optical
coupler and its approximated special case of the asymmetric nonlinear
optical coupler can operate under two conditions: spontaneous and
stimulated. In the spontaneous (stimulated) case, initially, i.e.,
at $t=0,$ there is no photon (non-zero number of photons) in the
second-harmonic mode of the system, whereas average photon numbers
in the other modes are non-zero. 

Following earlier works of some of the present authors \cite{Rechacek-2001-zrno-coupler,thun2002zeno-raman,chi2-chi1-spie,chi2-chi2,rehacek2000zeno-coupler},
the effect of the presence of the probe mode on the photon statistics
of the second harmonic mode of the system is investigated using Zeno
parameter ($\Delta N_{Z}$), which is defined as 
\begin{equation}
\Delta N_{Z}=\left\langle N_{X}\left(z\right)\right\rangle -\left\langle N_{X}\left(z\right)\right\rangle _{k=0}.\label{eq:zeno-parameter}
\end{equation}

The Zeno parameter is a measure of the effect caused on the evolution
of the photon statistics of the system (obtained for mode $X$) due
to its interaction with the probe. It can be inferred from Eq. (\ref{eq:zeno-parameter})
that the negative values of the Zeno parameter signify the continuous
measurement via the probe inhibited the evolution of mode $X$ by
decreasing the photon generation in that particular mode, which demonstrate
presence of the quantum Zeno effect. On the other hand, the positive
values of the Zeno parameter correspond to enhancement of the photon
generation due to coupling with auxiliary mode in the probe. This
is the signature of the presence of quantum anti-Zeno effect. 

\begin{figure}[H]
\includegraphics[angle=0,scale=0.3]{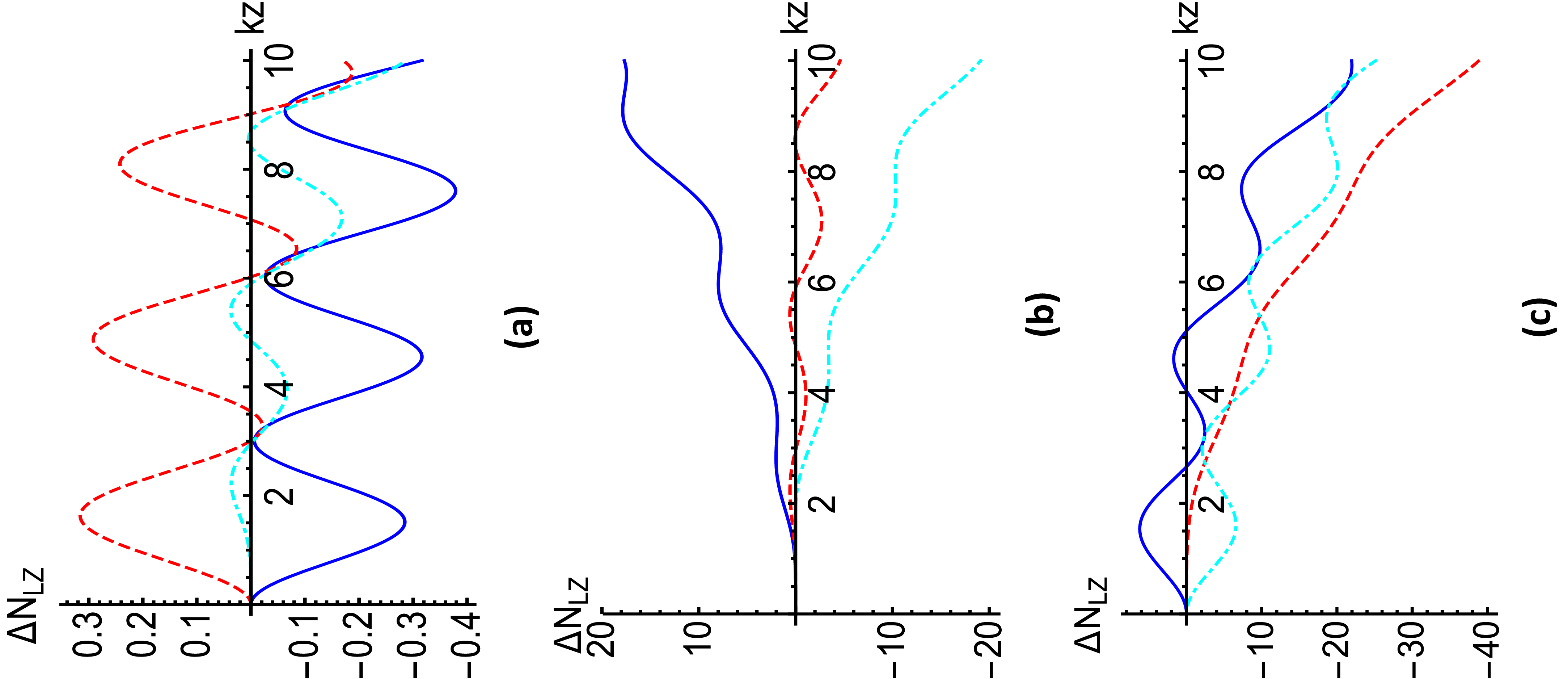}
\protect\caption{\label{fig:Linear-Zeno-z}(Color online) (a) The variation in linear
Zeno parameter with rescaled interaction length for $\frac{\Gamma_{b}}{k}=10^{-2},\,\frac{\Delta k_{b}}{k}=10^{-3}$
with $\alpha=5,\,\beta=3,$ and $\delta=1$ and -1 in stimulated case
(smooth and dashed lines) and $\delta=0$ in spontaneous case (dot-dashed
line). (b) In spontaneous case, $\alpha=10$ with $\beta=3,6,$ and
7 for smooth, dashed and dot-dashed lines, respectively. (c) $\alpha=10,\,\beta=8,$
and $\delta=-4,\,0,$ and 4 for smooth, dashed and dot-dashed lines,
respectively.}
\end{figure}

For the system of our interest, the symmetric nonlinear optical coupler,
using the analytic expression of the photon number operator in Eq.
(\ref{eq:nb2}), the Zeno parameter can be calculated for second harmonic
mode of the system waveguide as 

\begin{equation}
\begin{array}{lcl}
\Delta N_{NZ} & = & \left(\left|l_{2}\right|^{2}-\left|p_{2}\right|^{2}\right)\left|\beta\right|^{4}+\left|l_{3}\right|^{2}\left|\alpha\right|^{2}\left|\beta\right|^{2}\\
 & + & \left|l_{4}\right|^{2}\left|\alpha\right|^{4}+\left[\left(l_{2}-p_{2}\right)\beta^{2}\delta^{*}+l_{3}\alpha\beta\delta^{*}+l_{4}\alpha^{2}\delta^{*}\right.\\
 & + & l_{2}^{*}l_{3}\left|\beta\right|^{2}\alpha\beta^{*}+l_{2}^{*}l_{4}\alpha^{2}\beta^{*2}+l_{3}^{*}l_{4}\left|\alpha\right|^{2}\alpha\beta^{*}\\
 & + & \left(l_{5}-p_{5}\right)\left|\beta\right|^{2}\left|\delta\right|^{2}+\left(l_{6}-p_{6}\right)\left|\delta\right|^{2}+l_{7}\left|\delta\right|^{2}\alpha\beta^{*}\\
 & + & l_{8}\left|\delta\right|^{2}\alpha^{*}\beta+l_{9}\left|\alpha\right|^{2}\left|\delta\right|^{2}+l_{10}\alpha^{*}\beta\gamma\delta^{*}\\
 & + & l_{11}\left|\alpha\right|^{2}\gamma\delta^{*}+l_{12}\gamma\delta^{*}+l_{13}\left|\beta\right|^{2}\gamma\delta^{*}\\
 & + & \left.l_{14}\alpha\beta^{*}\gamma\delta^{*}+{\rm c.c.}\right],
\end{array}\label{eq:NL-zeno-parameter}
\end{equation}
where 
\begin{equation}
\begin{array}{lcl}
p_{2} & = & -\frac{\Gamma_{b}G_{-}^{*}}{\Delta k_{b}},\\
p_{5} & = & 2p_{6}=-\frac{4\left|\Gamma_{b}\right|^{2}\left(G_{-}^{*}+i\Delta k_{b}z\right)}{\left(\Delta k_{b}\right)^{2}}.
\end{array}\label{eq:atk0}
\end{equation}
Here, $p_{i}$s are obtained by taking $k=0$ in corresponding $l_{i}$s
in Eq. (\ref{eq:coefficient-l}) in Appendix A. All the remaining
$p_{i}'$s vanishes in the absence of the probe. The subscript $NZ$
in the Zeno parameter corresponds to the physical situation where
a nonlinear probe is used, i.e., ``nonlinear Zeno'' effect is investigated.
Thus $\Delta N_{NZ}$ can be referred to as the nonlinear Zeno parameter.
Similarly $\Delta N_{LZ}$ will denote linear Zeno parameter, i.e.,
Zeno parameter for a physical situation where linear probe is used.

It is easy to obtain Zeno parameter for the spontaneous case. Specifically,
in the spontaneous case, i.e., in absence of any photon in the second
harmonic mode of the system at $t=0$ (or considering $\delta=0$
at $t=0)$, the analytic expression of the nonlinear Zeno parameter
can be obtained from Eq. (\ref{eq:NL-zeno-parameter}) by keeping
the $\delta$ independent terms as 
\begin{equation}
\begin{array}{lcl}
\left(\Delta N_{NZ}\right)_{\delta=0} & = & \left(\left|l_{2}\right|^{2}-\left|p_{2}\right|^{2}\right)\left|\beta\right|^{4}+\left|l_{3}\right|^{2}\left|\alpha\right|^{2}\left|\beta\right|^{2}\\
 & + & \left|l_{4}\right|^{2}\left|\alpha\right|^{4}+\left[l_{2}^{*}l_{3}\left|\beta\right|^{2}\alpha\beta^{*}\right.\\
 & + & \left.l_{2}^{*}l_{4}\alpha^{2}\beta^{*2}+l_{3}^{*}l_{4}\left|\alpha\right|^{2}\alpha\beta^{*}+{\rm c.c.}\right].
\end{array}\label{eq:spon-nonlin-Z-par}
\end{equation}

In Section \ref{sec:System-and-solution}, we have already mentioned
that the momentum operator for an asymmetric nonlinear optical coupler
($\chi^{\left(2\right)}-\chi^{\left(1\right)})$ can be obtained by
just neglecting the nonlinear coupling term in one of the nonlinear
waveguides present in the symmetric nonlinear coupler ($\chi^{\left(2\right)}-\chi^{\left(2\right)}$)
studied here. Thus, we may consider the probe in the nonlinear Zeno
parameter obtained in Eq. (\ref{eq:NL-zeno-parameter}) to be linear
by taking $\Gamma_{a}=0$. This is how we can obtain the expression
for linear Zeno parameter. Thus, the Zeno parameter of the asymmetric
nonlinear optical coupler characterized by Eq. (\ref{eq:Ham-asymmetric})
can be obtained as 
\begin{equation}
\begin{array}{lcl}
\Delta N_{LZ} & = & \left(\left|l_{2}\right|^{2}-\left|p_{2}\right|^{2}\right)\left|\beta\right|^{4}+\left|l_{3}\right|^{2}\left|\alpha\right|^{2}\left|\beta\right|^{2}\\
 & + & \left|l_{4}\right|^{2}\left|\alpha\right|^{4}+\left[\left(l_{2}-p_{2}\right)\beta^{2}\delta^{*}+l_{3}\alpha\beta\delta^{*}+l_{4}\alpha^{2}\delta^{*}\right.\\
 & + & l_{2}^{*}l_{3}\left|\beta\right|^{2}\alpha\beta^{*}+l_{2}^{*}l_{4}\alpha^{2}\beta^{*2}+l_{3}^{*}l_{4}\left|\alpha\right|^{2}\alpha\beta^{*}\\
 & + & \left(l_{5}-p_{5}\right)\left|\beta\right|^{2}\left|\delta\right|^{2}+\left(l_{6}-p_{6}\right)\left|\delta\right|^{2}+l_{7}\left|\delta\right|^{2}\alpha\beta^{*}\\
 & + & \left.l_{8}\left|\delta\right|^{2}\alpha^{*}\beta+l_{9}\left|\alpha\right|^{2}\left|\delta\right|^{2}+{\rm c.c.}\right].
\end{array}\label{eq:L-zeno-parameter}
\end{equation}
Further, if we neglect all the terms beyond linear power in nonlinear
coupling constant $\Gamma_{b}$ in Eq. (\ref{eq:L-zeno-parameter}),
we find the result obtained here matches exactly with the result reported
in Ref. \cite{chi2-chi1-spie}. Interestingly, the analytic expressions
of the nonlinear and linear Zeno parameters have the same expressions
in the spontaneous case. It can also be checked that the expression
obtained in Eq. (\ref{eq:spon-nonlin-Z-par}) vanishes, if we neglect
all the terms beyond linear powers in the nonlinear coupling constant.
This is also consistent with the earlier result \cite{chi2-chi1-spie}.

\begin{figure}[H]
\includegraphics[angle=0,scale=0.35]{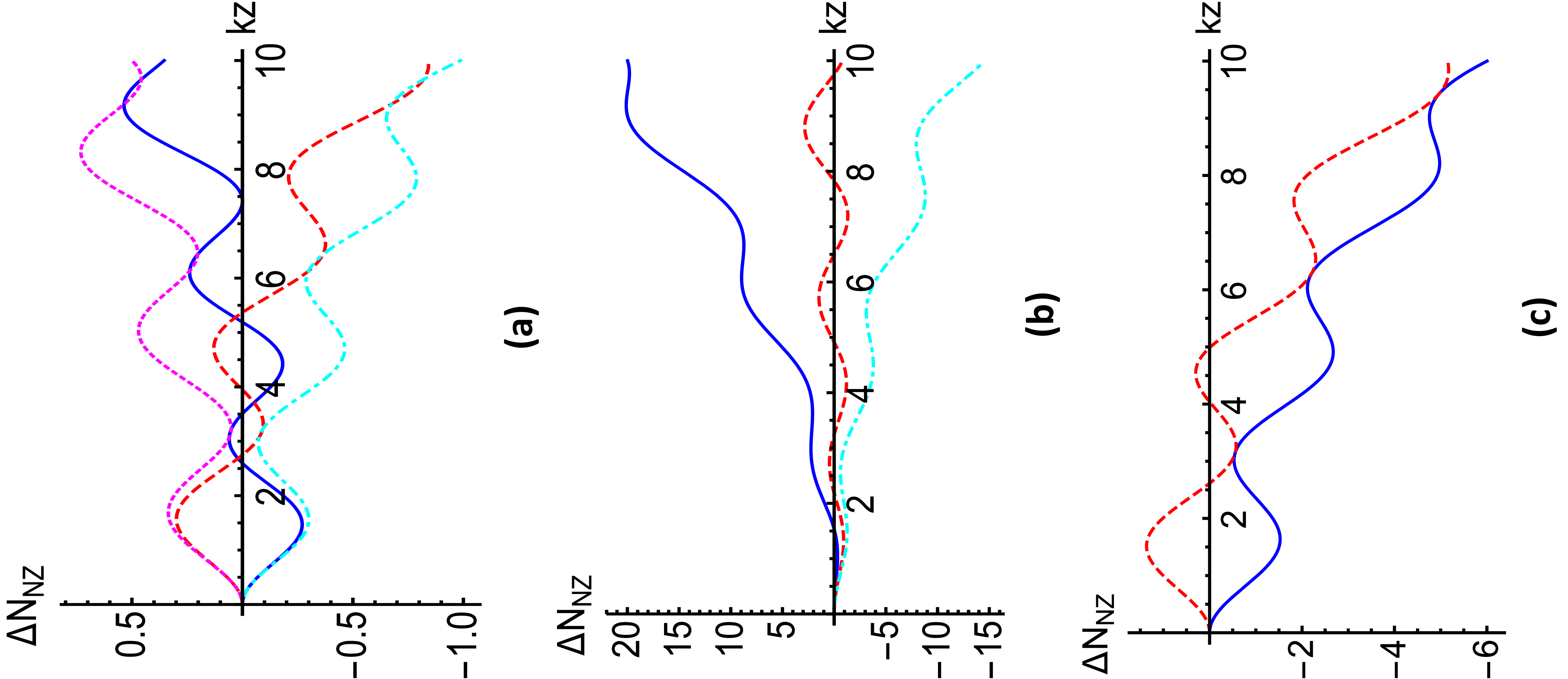}
\protect\caption{\label{fig:Nonlinear-z}(Color online) (a) Nonlinear Zeno parameter
with $\frac{\Gamma_{a}}{k}=\frac{\Gamma_{b}}{k}=10^{-2},\,\frac{\Delta k_{a}}{k}=1.1\times10^{-3},\,\frac{\Delta k_{b}}{k}=1.1\times10^{-3}$
with $\alpha=5,\,\beta=3,$ and $\gamma=2,\,\delta=1$ and -1 in stimulated
case (smooth and dashed lines) and $\gamma=-2,\,\delta=1$ and -1
in dot-dashed and dotted lines. (b) $\alpha=10,\,\gamma=3,\,\delta=1$
with $\beta=3,6,$ and 7 for smooth, dashed and dot-dashed lines,
respectively. (c) shows the change in nonlinear Zeno parameter with
changing the phase of $\alpha$ or $\beta$ by an amount of $\pi$\textcolor{red}{{}
}for $\alpha=\beta=6,$ and $\gamma=3,\,\delta=2$. It is also observed
that the change in phase of $\alpha$ is equivalent to change in phase
of $\beta$.}
\end{figure}

\subsection{Variation of Zeno parameter with different variables}

The analytic expressions obtained for both nonlinear and linear Zeno
parameters depend on various parameters, such as photon numbers and
phases of different field modes, linear and nonlinear coupling, interaction
length and phase mismatch between fundamental and second harmonic
modes in the nonlinear waveguides. However, in the spontaneous case,
system shows quantum anti-Zeno effect initially which eventually goes
towards quantum Zeno effect with increase in rescaled interaction
length. This behavior of the linear Zeno parameter in the spontaneous
case is further elaborated in Fig. \ref{fig:Linear-Zeno-z} b, where
it can be observed that as the number of photons in the linear mode
of the system waveguide becomes comparable to the photon numbers in
the probe mode, quantum Zeno effect is prominent. A similar effect
on photon numbers is observed even in the stimulated case in Fig.
\ref{fig:Linear-Zeno-z} c, where the transition to quantum Zeno effect
with increasing rescaled interaction length is more dominating than
in Fig. \ref{fig:Linear-Zeno-z} a. Further, in the stimulated case,
a transition between linear quantum Zeno and anti-Zeno effects can
be obtained by controlling the phase of second harmonic mode of the
system waveguide as illustrated in Fig. \ref{fig:Linear-Zeno-z} a.
However, with an increase in number of photons in fundamental mode
of the system waveguide shown in Fig. \ref{fig:Linear-Zeno-z} c,
this nature disappears gradually due to its dominant effect to tend
towards quantum Zeno effect.

To illustrate the variation of the nonlinear Zeno parameter with the
rescaled interaction length of the coupler in Fig. \ref{fig:Nonlinear-z},
we have considered specific values of all the remaining parameters.
As in the case of linear Zeno parameter (cf. Fig. \ref{fig:Linear-Zeno-z})
nonlinear Zeno parameter also shows dependence on the phases of both
second harmonic modes involved in the symmetric coupler. Specifically,
Fig. \ref{fig:Nonlinear-z} a illustrates that the change in phase
of the second harmonic mode of the system creates some changes in
the photon statistics which causes a transition between quantum Zeno
and anti-Zeno effects. This become more dominant with the change of
phase of second harmonic mode of the probe as well. Similarly, Fig.
\ref{fig:Nonlinear-z} b establishes an analogous fact for nonlinear
Zeno parameter as in Fig. \ref{fig:Linear-Zeno-z} b for linear Zeno
parameter, i.e., when the photon numbers in the linear modes of both
the waveguides are comparable then quantum Zeno effect prevails. Fig.
\ref{fig:Nonlinear-z} c shows that by changing the phase of $\alpha$
by $\pi$ (i.e., transforming $\alpha$ to $-\alpha$) has a similar
effect as changing the phase of $\beta$ by the same amount. Interestingly,
this kind of nature can be attributed to the symmetry present in the
symmetric coupler studied here.
\begin{widetext}

\begin{figure}[H]
\begin{centering}
\includegraphics[angle=-90,scale=0.9]{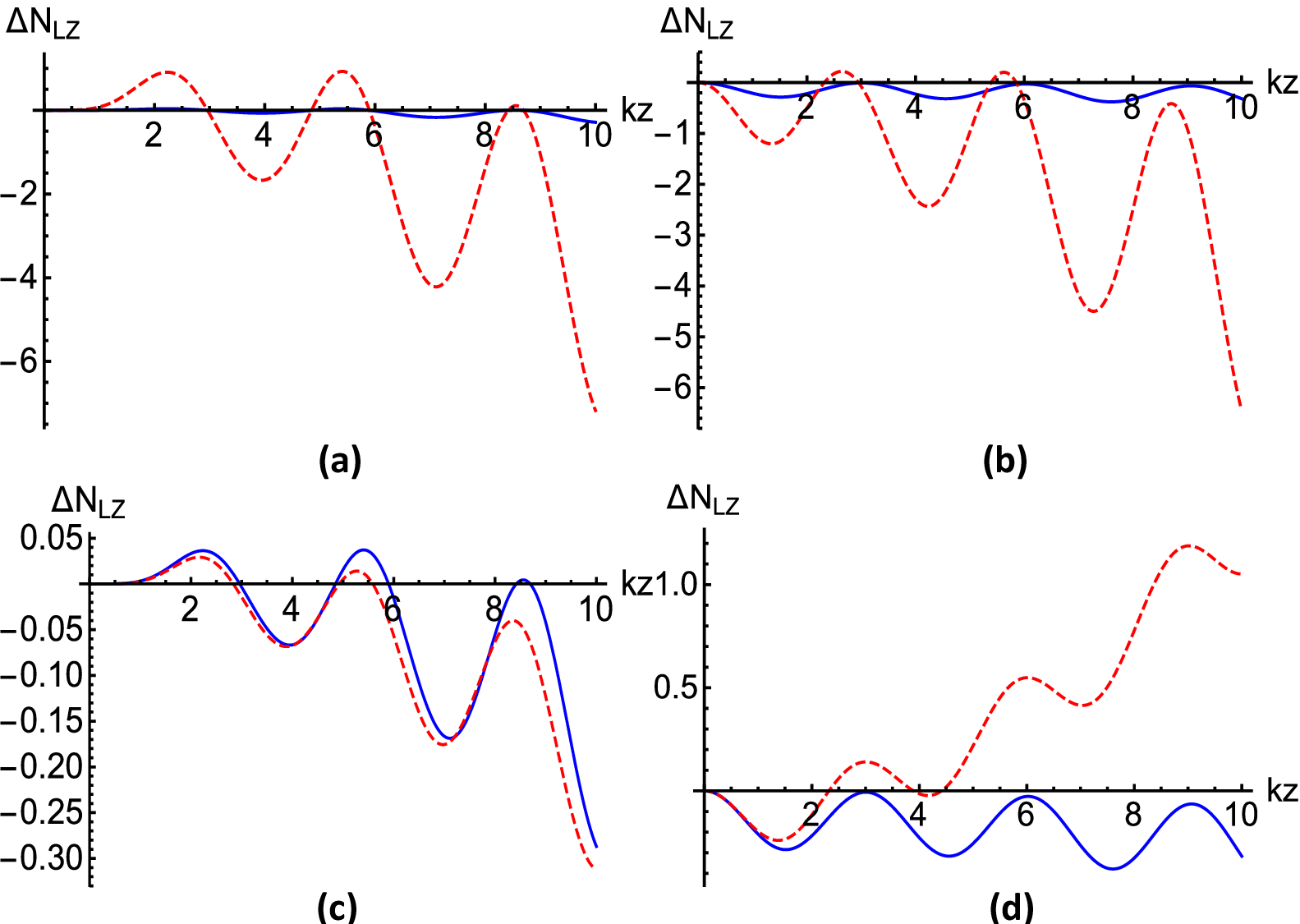}
\par\end{centering}
\protect\caption{\label{fig:Linear-Zeno-kz}(Color online) (a) Linear Zeno parameter
in the spontaneous case for $\frac{\Delta k_{b}}{k}=10^{-3}$ with
$\alpha=5,\,\beta=3$ for $\frac{\Gamma_{b}}{k}=10^{-2}\,\left(5\times10^{-2}\right)$
in smooth blue (dashed red) line. (b) A similar observation in the
stimulated case with $\delta=1$ and all the remaining values same
as (a). In (c) and (d), the effect of phase mismatch in the spontaneous
and stimulated (with $\delta=1$) cases of linear Zeno parameter is
shown, respectively. The remaining parameters are $\frac{\Gamma_{b}}{k}=10^{-2}$
with $\frac{\Delta k_{b}}{k}=10^{-3}\,\left(10^{-1}\right)$ in the
smooth blue (dashed red) line.}
\end{figure}

\end{widetext}

\begin{widetext}

\begin{figure}[H]
\begin{centering}
\includegraphics[angle=-90]{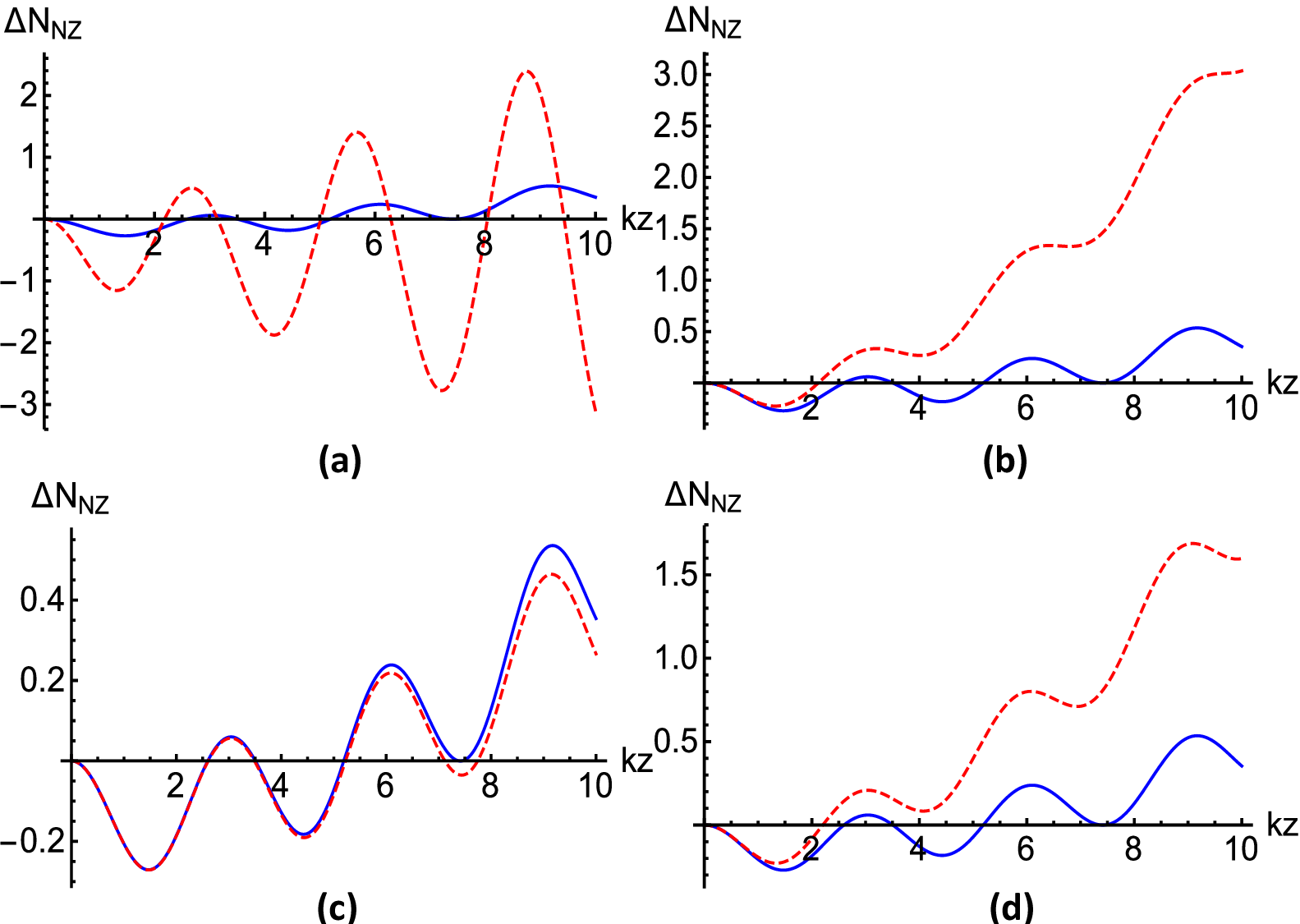}
\par\end{centering}

\protect\caption{\label{fig:Nonlinear-Zeno-kz}(Color online) The dependence of nonlinear
Zeno parameter on the nonlinear coupling constant is depicted in (a)
and (b). In (a) the nonlinear Zeno parameter is shown with rescaled
interaction length for $\frac{\Gamma_{a}}{k}=10^{-2},\,\frac{\Delta k_{a}}{k}=1.1\times10^{-3},\,\frac{\Delta k_{b}}{k}=10^{-3}$
with $\alpha=5,\,\beta=3,\gamma=2,$ and $\delta=1.$ The smooth (blue)
and dashed (red) lines correspond to $\frac{\Gamma_{b}}{k}=10^{-2}$
and $\frac{\Gamma_{b}}{k}=5\times10^{-2},$ respectively. Similarly,
in (b) the smooth (blue) and dashed (red) lines correspond to $\frac{\Gamma_{a}}{k}=10^{-2}$
and $\frac{\Gamma_{a}}{k}=5\times10^{-2},$ respectively with $\frac{\Gamma_{b}}{k}=10^{-2}$
and all the remaining values as in (a). In (c) the nonlinear Zeno
parameter is shown in the smooth blue (dashed red) line with $\frac{\Gamma_{a}}{k}=\frac{\Gamma_{b}}{k}=10^{-2},\,\frac{\Delta k_{b}}{k}=10^{-3}$
for $\frac{\Delta k_{a}}{k}=1.1\times10^{-3}\,\left(1.1\times10^{-1}\right).$
Similarly, in (d) the nonlinear Zeno parameter is shown in the smooth
blue (dashed red) line with $\frac{\Delta k_{a}}{k}=1.1\times10^{-3}$
for $\frac{\Delta k_{b}}{k}=10^{-3}\,\left(10^{-1}\right).$}
\end{figure}

\end{widetext}

The explicit dependence of the linear and nonlinear Zeno parameters
on the remaining parameters, such as nonlinear coupling constants
and phase mismatches, of both system and probe waveguides is illustrated
in Figs. \ref{fig:Linear-Zeno-kz} and \ref{fig:Nonlinear-Zeno-kz},
respectively. Specifically, Figs. \ref{fig:Linear-Zeno-kz} a-b show
the variation in the linear Zeno parameter for two values of nonlinear
coupling constant of the system in spontaneous and stimulated cases,
respectively. A similar study is shown in Figs. \ref{fig:Nonlinear-Zeno-kz}
a-b for the nonlinear Zeno parameter with two values of nonlinear
coupling constants of the system and probe waveguides, respectively.
All the cases demonstrate that with increase in the nonlinear coupling
of the system a dominant oscillatory nature is observed. While increase
in the nonlinear coupling of the probe waveguide shows preference
for quantum anti-Zeno effect.

\begin{figure}[H]
\includegraphics[angle=0,scale=0.4]{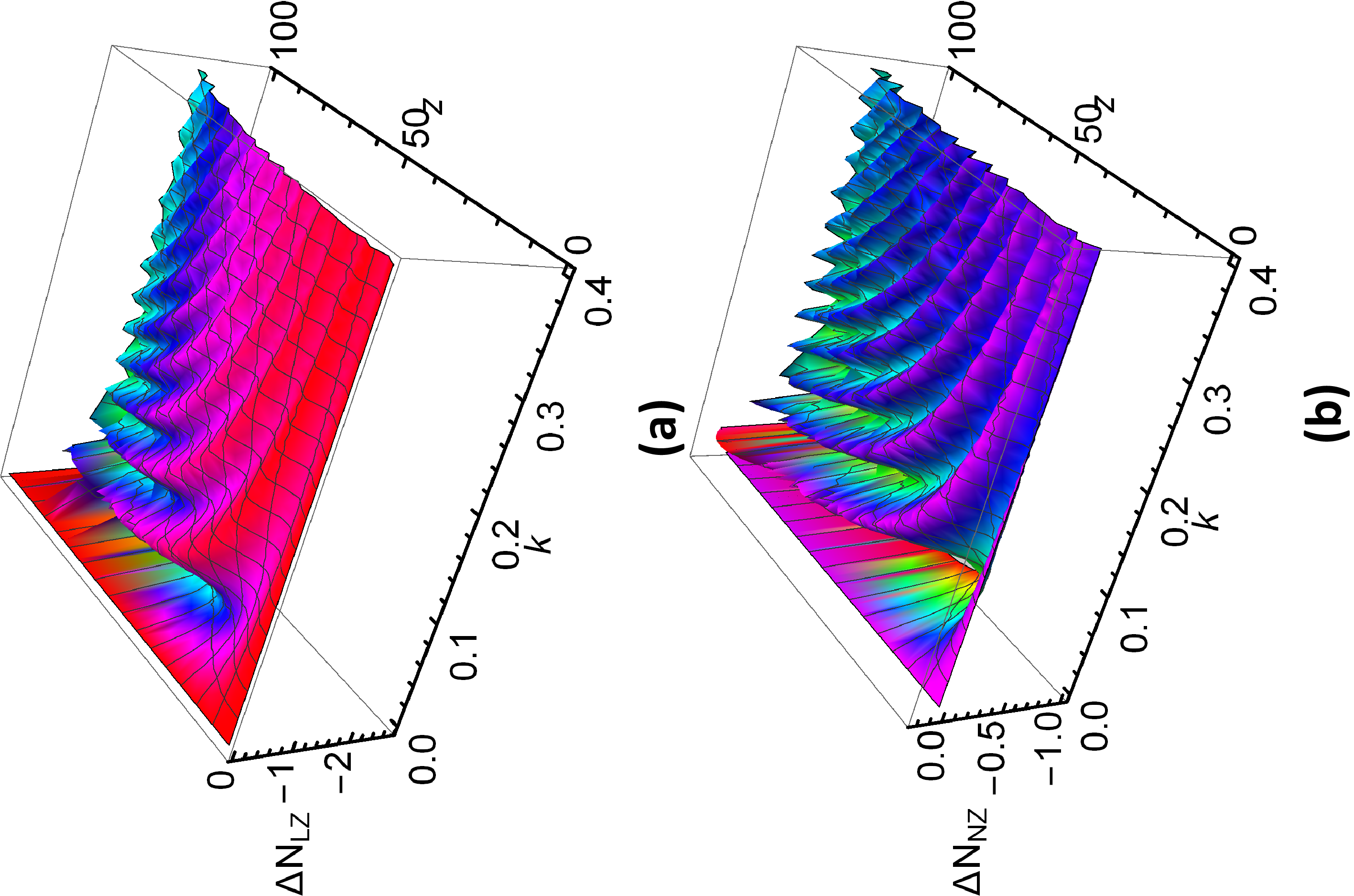}

\protect\caption{\label{fig:with-k}(Color online) The variation in (a) linear and
(b) nonlinear Zeno parameter with linear coupling constant and interaction
length are shown for $\frac{\Gamma_{a}}{k}=\frac{\Gamma_{b}}{k}=10^{-2},\,\frac{\Delta k_{a}}{k}=1.1\times10^{-3},\,\frac{\Delta k_{b}}{k}=10^{-3}$
with $\alpha=6,\,\beta=4,\,\gamma=2,$ and $\delta=1$. }
\end{figure}

The phase mismatch between fundamental and second harmonic modes of
the system (probe) waveguide has negligible effect on the linear (nonlinear)
Zeno parameter in the spontaneous (stimulated) case as depicted in
Fig. \ref{fig:Linear-Zeno-kz} c (\ref{fig:Nonlinear-Zeno-kz} c).
While a similar observation for linear (nonlinear) Zeno parameter
shown in Fig. \ref{fig:Linear-Zeno-kz} d (\ref{fig:Nonlinear-Zeno-kz}
d) for the stimulated case with phase mismatch in the system waveguide
exhibits a transition from quantum Zeno effect to quantum anti-Zeno
effect.
The dependence of both linear and nonlinear Zeno parameters on the
linear coupling can be observed with interaction length in Fig. \ref{fig:with-k}
a and b, respectively. With the particular choice of values for other
parameters, both Zeno parameters show quantum Zeno effect. However,
as observed in Fig. \ref{fig:Linear-Zeno-z} quantum anti-Zeno effect
can be illustrated here by just controlling the phase of the second
harmonic mode in the system. Though an increase in the effect of the
presence of the probe in the photon statistics of the second harmonic
mode with increasing interaction length and linear coupling can be
observed from the figure. This dominance of the effect of the probe
is oscillatory in nature and gives a ripple like structure in Fig.
\ref{fig:with-k}.

The effect of change in phase mismatch between fundamental and second
harmonic modes explored in Figs. \ref{fig:Linear-Zeno-kz} c-d and
\ref{fig:Nonlinear-Zeno-kz} c-d is further illustrated in Fig. \ref{fig:with-mismatch}. The phase mismatch between the fundamental and second harmonic modes
in the system waveguide has evident effect on the linear Zeno parameter
only in the small mismatch region for the spontaneous case (cf. Fig.
\ref{fig:with-mismatch} a). Whereas an increase in the initial number
of photons in the second harmonic mode of the system, i.e., in the
stimulated case, changes the photon statistics drastically and both
quantum Zeno and anti-Zeno effects, with continuous switching between
them, have been observed in Fig. \ref{fig:with-mismatch} b. The corresponding
plot for nonlinear Zeno parameter shows a quite similar behavior in
Fig. \ref{fig:with-mismatch} c with slight changes in the photon
statistics due to the presence of the second harmonic mode in the
probe. A similar study for the effect of phase mismatch between the
fundamental and second harmonic modes of the probe on nonlinear Zeno
parameter shows ample amount of variation only for small mismatch
and becomes almost constant for larger values of phase mismatch. These features of quantum Zeno and anti-Zeno effects can be further illustrated using contour plots as shown in Fig. \ref{fig:with-mismatch-cont}, where the values of different parameters are the same as those used in Fig. \ref{fig:with-mismatch}. The contour plots can be drawn to clearly show the regions of quantum Zeno and anti-Zeno effects (without referring to the magnitude of the Zeno parameter) as shown in Fig. \ref{fig:with-mismatch-cont} a and b, where the blue regions correspond to the quantum Zeno effect while the yellow regions correspond to the quantum anti-Zeno effect in the linear Zeno case. The contour plots can also be drawn to illustrate the depth of Zeno parameter for both the effects as illustrated in Fig. \ref{fig:with-mismatch-cont} c and d for nonlinear Zeno parameter.

Fig. \ref{fig:Linear-ab} demonstrates the nature of linear Zeno parameter
with changes in the number of photons in the linear modes of both
the waveguides. Here, it can be seen that with increase in photon
numbers in probe mode quantum anti-Zeno effect is preferred while
with increasing the intensity in the linear mode of system waveguide
it tends towards quantum Zeno effect.

\section{Conclusions \label{sec:Conclusions}}

\begin{figure}
\includegraphics[angle=0,scale=0.4]{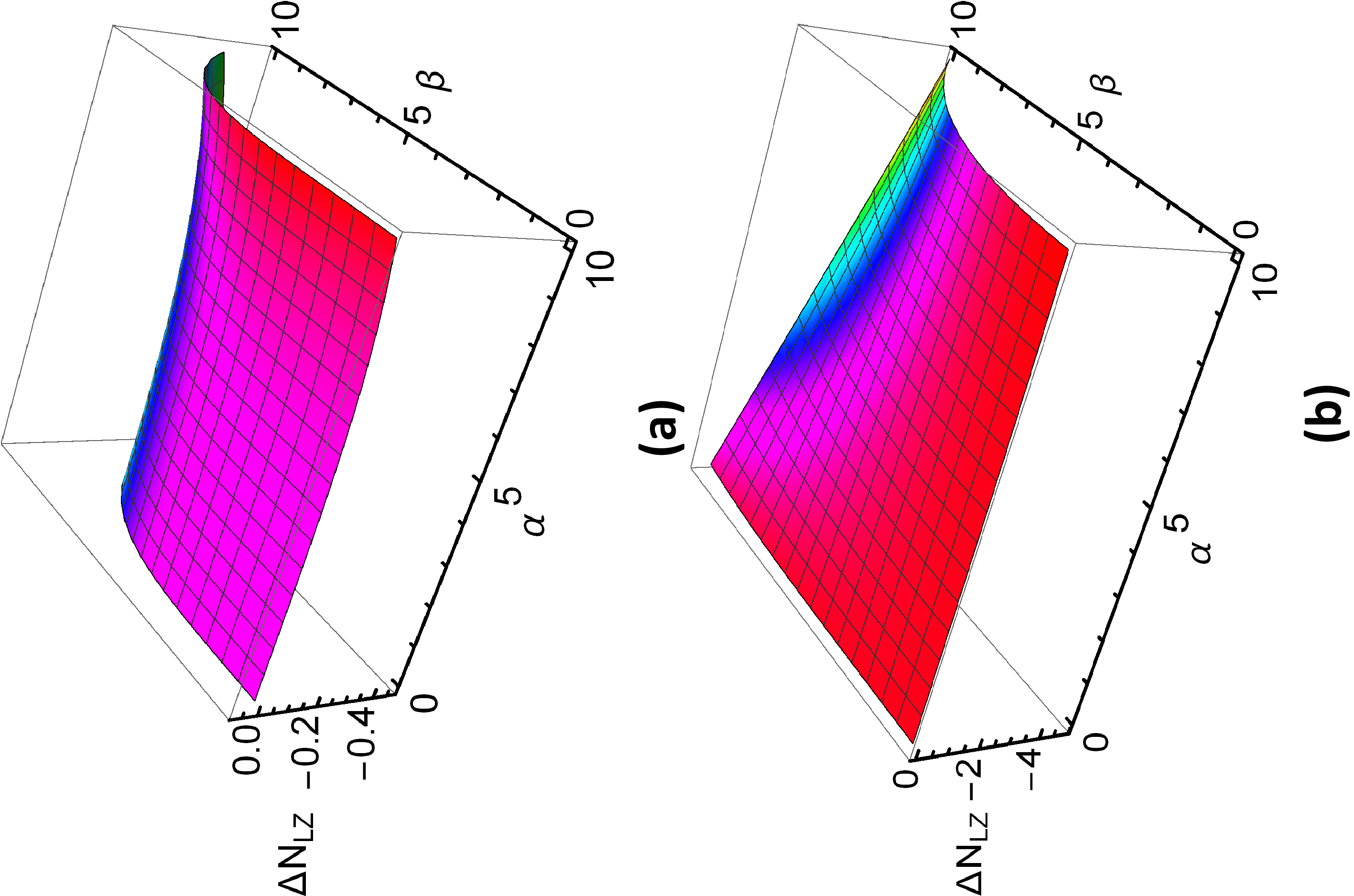}

\protect\caption{\label{fig:Linear-ab}(Color online) The variation in linear Zeno
parameter in (a) spontaneous and (b) stimulated cases with photon
numbers of linear modes in both the waveguides ($\alpha$ and $\beta$)
are shown for $\frac{\Gamma_{b}}{k}=10^{-2},\,\frac{\Delta k_{b}}{k}=10^{-3}$
with $\gamma=\frac{\alpha}{2}$ and $\delta=\frac{\beta}{3}$ after
rescaled interaction length $kz=1$. A similar behavior to the linear
Zeno parameter is observed for nonlinear Zeno parameter in the stimulated
case. }
\end{figure}

Linear and nonlinear quantum Zeno and anti-Zeno effects in a symmetric
and an asymmetric nonlinear optical couplers are rigorously investigated
in the present work. The investigation is performed using linear and
nonlinear Zeno parameters, which are introduced in this paper in analogy
with that of the Zeno parameter introduced in Ref. \cite{chi2-chi1-spie}.
Closed form analytic expressions for both linear and nonlinear Zeno
parameters are obtained here using Sen-Mandal perturbative method.
Subsequently, variation of the Zeno and anti-Zeno parameters with
respect to various quantities are investigated and the same is illustrated
in Figs. \ref{fig:Linear-Zeno-z}-\ref{fig:Linear-ab}. The investigation
led to several interesting observations. For example, we have observed
that the analytic expressions obtained for both linear and nonlinear
Zeno parameters are the same for the spontaneous case. Further, in
the spontaneous case, it is observed that the transition from the
quantum anti-Zeno effect to quantum Zeno effect can be achieved by
increasing the intensity of the radiation field in the linear mode
of the system waveguide (cf. Fig. \ref{fig:Linear-Zeno-z} b). Similarly,
a switching between the linear (nonlinear) quantum Zeno and anti-Zeno
effects is also observed in the stimulated case. However, it is observed
that this switching can be obtained just by controlling the phase
of the second harmonic mode in the system waveguide in the linear
case (cf. Figs. \ref{fig:Linear-Zeno-z} a and c) and by controlling
the phase of the nonlinear modes of both the waveguides (cf. Figs.
\ref{fig:Nonlinear-z} a and c). Here we may note that the change
in phase of the linear mode of the probe is equivalent to change in
phase of the linear mode of the nonlinear system waveguide. This kind
of nature can be attributed to the symmetry present in the system,
which is evident even in the system momentum operator (cf. Fig. \ref{fig:Nonlinear-z}
c). In fact, in general, we have observed that increase in the intensity
of the probe leads to increase the quantum anti-Zeno effect, while
with the increase in the intensity of the linear mode of the system
waveguide the quantum Zeno effect is more prominent (cf. Figs. \ref{fig:Linear-Zeno-z}
b and c and Fig. \ref{fig:Linear-ab}). 
\begin{widetext}

\begin{figure}[H]
\begin{centering}
\includegraphics[angle=-90,scale=0.75]{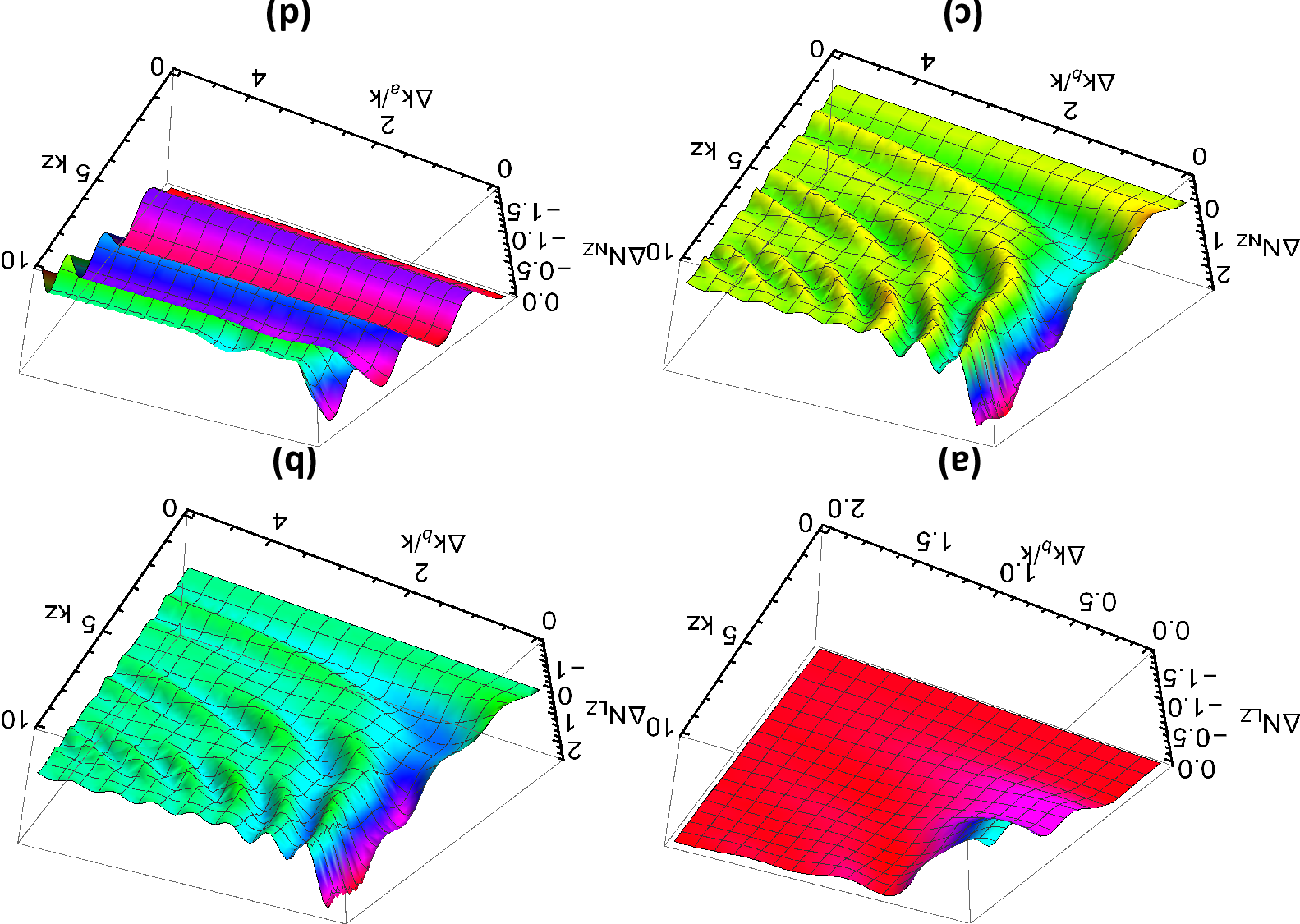}
\par\end{centering}

\protect\caption{\label{fig:with-mismatch}(Color online) The variation in linear Zeno
parameter with phase mismatch between fundamental and second harmonic
modes in system waveguide and rescaled interaction length in (a) spontaneous
and (b) stimulated cases are shown for $\frac{\Gamma_{b}}{k}=10^{-2}$
with $\alpha=6,\,\beta=4,$ and $\delta=0$ and 1 in (a) and (b),
respectively. (c) shows the dependence of the nonlinear Zeno parameter
on the phase mismatch between fundamental and second harmonic modes
in system waveguide and rescaled interaction length for $\frac{\Gamma_{a}}{k}=10^{-2},\,\frac{\Delta k_{a}}{k}=1.1\times10^{-3}$
and $\gamma=2,\,\delta=1$ with all remaining values same as (a) and
(b). In (d), the effect of phase mismatch between fundamental and
second harmonic modes in probe waveguide and rescaled interaction
length on the nonlinear Zeno parameter are shown for $\frac{\Delta k_{b}}{k}=10^{-3}$
with all the remaining values as (c).}
\end{figure}

\begin{figure}[H]
\begin{centering}
\includegraphics[angle=-90,scale=0.75]{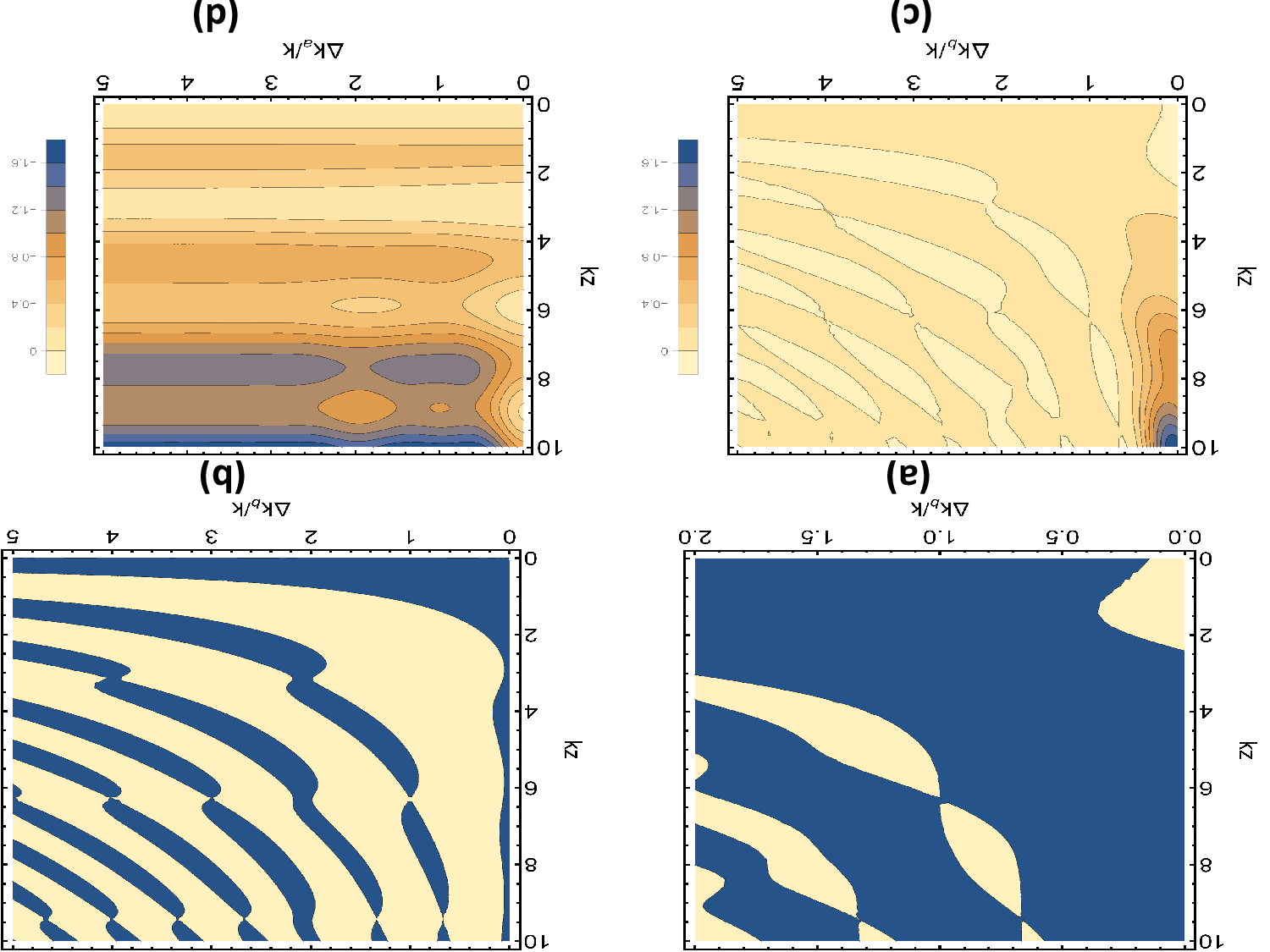} 
\par\end{centering}

\protect\caption{\label{fig:with-mismatch-cont}(Color online) Variation in linear and nonlinear Zeno parameters shown in three dimensional plots in Fig. \ref{fig:with-mismatch} (a)-(d) are illustrated via equivalent contour plots. Here, all the four contour plots corresponding to Fig. \ref{fig:with-mismatch} (a)-(d) are are obtained using the same parameters. In (a) and (b), the yellow regions illustrate the region for quantum anti-Zeno effect while the blue regions correspond to quantum Zeno effect. In (c) and (d), along with the regions of Zeno and anti-Zeno effects, variation of magnitude of Zeno parameters are also shown with different colors (see the color bars in right side of the figures).}
\end{figure}

\end{widetext}

For the smaller values of the linear coupling constant, a considerable
amount of variation in the photon number statistics is observed through
the linear Zeno parameter. This variation is observed to fade away
as ripples with increasing interaction length for higher values of
linear coupling constant (cf. Fig. \ref{fig:with-k} a). Similar but
more prominent nature is observed in the nonlinear case (cf. Fig.
\ref{fig:with-k} b). Similarly, we have observed that with the increase
in phase mismatch between fundamental and second harmonic modes in
the system waveguide, a transition from quantum Zeno effect to quantum
anti-Zeno effect occurs. The change in photon number statistics of
the nonlinear waveguide is more prominent in the stimulated case compared
to that in the spontaneous case of linear and nonlinear Zeno parameters,
respectively (cf. Fig. \ref{fig:with-mismatch}).\textcolor{blue}{{} }

In brief, possibility of observing Zeno and anti-Zeno effects is rigorously
investigated in symmetric and asymmetric nonlinear optical couplers,
which are experimentally realizable at ease. For completely quantum
description of the primary physical system (i.e., symmetric nonlinear
optical coupler), appropriate use of a perturbative technique which
is known to perform better than short-length method, reducibility
of the results obtained for symmetric nonlinear optical coupler to
that of asymmetric nonlinear optical coupler, easy experimental realizability
of the physical systems, etc., provide an edge to this work over the
existing works on Zeno effects in optical coupler, where usually the
use of complete quantum description is circumvented by considering
one or more modes as strong and/or short length method is used to
reduce computational difficulty. The approach adopted here is also
very general and can be easily extended to the study of other optical
couplers and other quantum optical systems having the similar structure
of momentum operators or Hamiltonian as is used here. We conclude
the work with an expectation that the experimentalists will find this
work interesting for an experimental verification and it could be
possible to find its applicability in some of the recently proposed
Zeno-effect-based schemes for quantum computation and communication.

\appendix

\section*{Appendix A: Solution of Heisenberg's equations of motion \label{sec:A}}

\setcounter{equation}{0} \renewcommand{\theequation}{A.\arabic{equation}} 
The solution of the momentum operator given in Eq. (\ref{eq:Ham-symmetric})
using Sen-Mandal perturbative approach can be obtained once we write
the Heisenberg's equations of motion for all the field modes involved,
which are obtained as

\begin{equation}
\begin{array}{lcl}
\frac{da_{1}}{dz} & = & ik^{*}b_{1}+2i\Gamma_{a}^{*}a_{1}^{\dagger}a_{2}\exp\left(-i\Delta k_{a}z\right),\\
\frac{db_{1}}{dz} & = & ika_{1}+2i\Gamma_{b}^{*}b_{1}^{\dagger}b_{2}\exp\left(-i\Delta k_{b}z\right),\\
\frac{da_{2}}{dz} & = & i\Gamma_{a}a_{1}^{2}\exp\left(i\Delta k_{a}z\right),\\
\frac{db_{2}}{dz} & = & i\Gamma_{b}b_{1}^{2}\exp\left(i\Delta k_{b}z\right).
\end{array}\label{eq:Heisenberg's-eqs}
\end{equation}
Now, the evolution the all the field modes can be assumed up to quadratic
terms in nonlinear coupling constants $\Gamma_{i}$ in the form 

\begin{widetext}

\begin{equation}
\begin{array}{lcl}
a_{1}(z) & = & f_{1}a_{1}(0)+f_{2}b_{1}(0)+f_{3}a_{1}^{\dagger}(0)a_{2}(0)+f_{4}a_{2}(0)b_{1}^{\dagger}(0)+f_{5}b_{1}^{\dagger}(0)b_{2}(0)+f_{6}a_{1}^{\dagger}(0)b_{2}(0)\\
 & + & f_{7}a_{1}(0)a_{2}^{\dagger}(0)a_{2}(0)+f_{8}a_{1}^{\dagger}(0)a_{1}^{2}(0)+f_{9}a_{2}^{\dagger}(0)a_{2}(0)b_{1}(0)+f_{10}a_{1}^{\dagger}(0)a_{1}(0)b_{1}(0)\\
 & + & f_{11}a_{1}(0)b_{1}^{\dagger}(0)b_{1}(0)+f_{12}a_{1}^{\dagger}(0)b_{1}^{2}(0)+f_{13}a_{1}^{2}(0)b_{1}^{\dagger}(0)+f_{14}b_{1}^{\dagger}(0)b_{1}^{2}(0)\\
 & + & f_{15}a_{2}(0)b_{1}(0)b_{2}^{\dagger}(0)+f_{16}a_{1}(0)a_{2}(0)b_{2}^{\dagger}(0)+f_{17}a_{1}(0)a_{2}^{\dagger}(0)b_{2}(0)\\
 & + & f_{18}a_{2}^{\dagger}(0)b_{1}(0)b_{2}(0)+f_{19}b_{1}(0)b_{2}^{\dagger}(0)b_{2}(0)+f_{20}b_{1}^{\dagger}(0)b_{1}^{2}(0)+f_{21}a_{1}^{\dagger}(0)b_{1}^{2}(0)\\
 & + & f_{22}a_{1}(0)b_{2}^{\dagger}(0)b_{2}(0)+f_{23}a_{1}(0)b_{1}^{\dagger}(0)b_{1}(0)+f_{24}a_{1}^{\dagger}(0)a_{1}(0)b_{1}(0)+f_{25}a_{1}^{2}(0)b_{1}^{\dagger}(0)\\
 & + & f_{26}a_{1}^{\dagger}(0)a_{1}^{2}(0),\\
b_{1}(z) & = & g_{1}a_{1}(0)+g_{2}b_{1}(0)+g_{3}a_{1}^{\dagger}(0)a_{2}(0)+g_{4}a_{2}(0)b_{1}^{\dagger}(0)+g_{5}b_{1}^{\dagger}(0)b_{2}(0)+g_{6}a_{1}^{\dagger}(0)b_{2}(0)\\
 & + & g_{7}a_{1}(0)a_{2}^{\dagger}(0)a_{2}(0)+g_{8}a_{1}^{\dagger}(0)a_{1}^{2}(0)+g_{9}a_{2}^{\dagger}(0)a_{2}(0)b_{1}(0)+g_{10}a_{1}^{\dagger}(0)a_{1}(0)b_{1}(0)\\
 & + & g_{11}a_{1}(0)b_{1}^{\dagger}(0)b_{1}(0)+g_{12}a_{1}^{\dagger}(0)b_{1}^{2}(0)+g_{13}a_{1}^{2}(0)b_{1}^{\dagger}(0)+g_{14}b_{1}^{\dagger}(0)b_{1}^{2}(0)\\
 & + & g_{15}a_{2}(0)b_{1}(0)b_{2}^{\dagger}(0)+g_{16}a_{1}(0)a_{2}(0)b_{2}^{\dagger}(0)+g_{17}a_{1}(0)a_{2}^{\dagger}(0)b_{2}(0)\\
 & + & g_{18}a_{2}^{\dagger}(0)b_{1}(0)b_{2}(0)+g_{19}b_{1}(0)b_{2}^{\dagger}(0)b_{2}(0)+g_{20}b_{1}^{\dagger}(0)b_{1}^{2}(0)+g_{21}a_{1}^{\dagger}(0)b_{1}^{2}(0)\\
 & + & g_{22}a_{1}(0)b_{2}^{\dagger}(0)b_{2}(0)+g_{23}a_{1}(0)b_{1}^{\dagger}(0)b_{1}(0)+g_{24}a_{1}^{\dagger}(0)a_{1}(0)b_{1}(0)+g_{25}a_{1}^{2}(0)b_{1}^{\dagger}(0)\\
 & + & g_{26}a_{1}^{\dagger}(0)a_{1}^{2}(0),\\
a_{2}(z) & = & h_{1}a_{2}(0)+h_{2}a_{1}^{2}(0)+h_{3}b_{1}(0)a_{1}(0)+h_{4}b_{1}^{2}(0)+h_{5}a_{1}^{\dagger}(0)a_{1}(0)a_{2}(0)+h_{6}a_{2}(0)\\
 & + & h_{7}a_{1}^{\dagger}(0)a_{2}(0)b_{1}(0)+h_{8}a_{1}(0)a_{2}(0)b_{1}^{\dagger}(0)+h_{9}a_{2}(0)b_{1}^{\dagger}(0)b_{1}(0)+h_{10}a_{1}(0)b_{1}^{\dagger}(0)b_{2}(0)\\
 & + & h_{11}b_{1}^{\dagger}(0)b_{1}(0)b_{2}(0)+h_{12}b_{2}(0)+h_{13}a_{1}^{\dagger}(0)a_{1}(0)b_{2}(0)+h_{14}a_{1}^{\dagger}(0)b_{1}(0)b_{2}(0),\\
b_{2}(z) & = & l_{1}b_{2}(0)+l_{2}b_{1}^{2}(0)+l_{3}b_{1}(0)a_{1}(0)+l_{4}a_{1}^{2}(0)+l_{5}b_{1}^{\dagger}(0)b_{1}(0)b_{2}(0)+l_{6}b_{2}(0)\\
 & + & l_{7}a_{1}(0)b_{1}^{\dagger}(0)b_{2}(0)+l_{8}a_{1}^{\dagger}(0)b_{1}(0)b_{2}(0)+l_{9}a_{1}^{\dagger}(0)a_{1}(0)b_{2}(0)+l_{10}a_{1}^{\dagger}(0)a_{2}(0)b_{1}(0)\\
 & + & l_{11}a_{1}^{\dagger}(0)a_{1}(0)a_{2}(0)+l_{12}a_{2}(0)+l_{13}a_{2}(0)b_{1}^{\dagger}(0)b_{1}(0)+l_{14}a_{1}(0)a_{2}(0)b_{1}^{\dagger}(0).
\end{array}\label{eq:assumed-sol}
\end{equation}

\end{widetext}

All the $f_{i},\,g_{i},\,h_{i},$ and $l_{i}$ can be obtained using
the assumed solution (\ref{eq:assumed-sol}) for different field modes
in the coupled differential equations in Eq. (\ref{eq:Heisenberg's-eqs})
with the boundary conditions for all $F_{1}\left(z=0\right)=1,$ where
$F\in\left\{ f,g,h,l\right\} $. The closed form analytic solutions
given in Eq. (\ref{eq:assumed-sol}) contain various coefficients,
for example, 

\[
\begin{array}{lcl}
l_{1} & = & 1,\\
l_{2} & = & -\frac{\Gamma_{b}G_{b-}^{*}}{2\Delta k_{b}}+\frac{iC_{b}}{2}\left[2|k|\left(G_{b+}^{*}-1\right)\sin2|k|z-i\Delta k_{b}\right.\\
 & \times & \left.\left(1-\left(G_{b+}^{*}-1\right)\cos2|k|z\right)\right],\\
l_{3} & = & \frac{-C_{b}|k|\left[i\Delta k_{b}\left(G_{b+}^{*}-1\right)\sin2|k|z+2|k|\left(1-\left(G_{b+}^{*}-1\right)\cos2|k|z\right)\right]}{k^{*}},\\
l_{4} & = & \frac{\Gamma_{b}|k|^{2}G_{b-}^{*}}{2k^{*^{2}}\Delta k_{b}}+\frac{iC_{b}|k|^{2}}{2k^{*^{2}}}\left[2|k|\left(G_{b+}^{*}-1\right)\sin2|k|z-i\Delta k_{b}\right.\\
 & \times & \left.\left(1-\left(G_{b+}^{*}-1\right)\cos2|k|z\right)\right],\\
l_{5} & = & \frac{\left|C_{b}\right|^{2}}{|k|\Delta k_{b}^{2}}\left[-16|k|^{5}\left(G_{b-}^{*}+i\Delta k_{b}z\right)-8i|k|^{4}\Delta k_{b}G_{b-}^{*}\right.\\
 & \times & \sin2|k|z+6i|k|^{2}\Delta k_{b}^{3}G_{b-}^{*}\sin2|k|z-i\Delta k_{b}^{5}\sin2|k|z\\
 & + & 4|k|^{3}\Delta k_{b}^{2}\left(\cos2|k|z-1+3G_{b-}^{*}+3i\Delta k_{b}z\right)+\Delta k_{b}^{4}\\
 & \times & \left.|k|\left(\left(1-2G_{b-}^{*}\right)\cos2|k|z-1-2G_{b-}^{*}-2i\Delta k_{b}z\right)\right],\\
l_{6} & = & \frac{\left|C_{b}\right|^{2}}{\Delta k_{b}^{2}}\left[-16|k|^{4}\left(G_{b-}^{*}+i\Delta k_{b}z\right)-4i|k|\Delta k_{b}^{3}\right.\\
 & \times & \left(G_{b+}^{*}-1\right)\sin2|k|z+4|k|^{2}\Delta k_{b}^{2}\\
 & \times & \left(\cos2|k|z\left(G_{b+}^{*}-1\right)+2G_{b-}^{*}-1+3i\Delta k_{b}z\right)\\
 & + & \left.\Delta k_{b}^{4}\left(\cos2|k|z\left(G_{b+}^{*}-1\right)-1-G_{b-}^{*}-2i\Delta k_{b}z\right)\right],\\
l_{7} & = & \frac{\left|C_{b}\right|^{2}}{k^{*}\Delta k_{b}}\left[2\Delta k_{b}^{4}\sin^{2}|k|z+4i|k|^{3}\Delta k_{b}\sin2|k|z\right.\\
 & - & i|k|\Delta k_{b}^{3}\left(2G_{b-}^{*}-1\right)\sin2|k|z\\
 & + & 8|k|^{4}\left(-G_{b-}^{*}\cos2|k|z+G_{b-}^{*}-i\Delta k_{b}z\right)\\
 & + & \left.2|k|^{2}\Delta k_{b}^{2}\left(3G_{b-}^{*}\cos2|k|z-G_{b-}^{*}+i\Delta k_{b}z\right)\right],
\end{array}
\]

\[
 \begin{array}{lcl}
l_{8} & = & -\frac{\left|C_{b}\right|^{2}}{k\Delta k_{b}}\left[2\Delta k_{b}^{4}\sin^{2}|k|z-i|k|\Delta k_{b}^{3}\sin2|k|z\right.\\
 & + & 4i|k|^{3}\Delta k_{b}\left(2G_{b-}^{*}-1\right)\sin2|k|z\\
 & + & 8|k|^{4}\left(-G_{b-}^{*}\cos2|k|z+G_{b-}^{*}+i\Delta k_{b}z\right)+2|k|^{2}\\
 & \times & \left.\Delta k_{b}^{2}\left(\left(G_{b+}^{*}+2\right)\cos2|k|z-G_{b-}^{*}-4-i\Delta k_{b}z\right)\right],\\
l_{9} & = & \frac{\left|C_{b}\right|^{2}}{|k|\Delta k_{b}^{2}}\left[-16|k|^{5}\left(G_{b-}^{*}+i\Delta k_{b}z\right)+\left\{ 8i|k|^{4}\Delta k_{b}G_{b-}^{*}\right.\right.\\
 & - & \left.2i|k|^{2}\Delta k_{b}^{3}\left(G_{b+}^{*}+2\right)+i\Delta k_{b}^{5}\right\} \sin2|k|z+4|k|^{3}\\
 & \times & \left(\left(1-2G_{b-}^{*}\right)\cos2|k|z-1+G_{b-}^{*}+3i\Delta k_{b}z\right)\\
 & \times & \left.\Delta k_{b}^{2}+|k|\Delta k_{b}^{4}\left(\cos2|k|z-1-2i\Delta k_{b}z\right)\right],\\
l_{10} & = & \frac{2C_{ab}|k|\left(G_{ab+}-1\right)}{k^{*}}\left[2i|k|^{2}\Delta k_{b}\Delta k_{ab}\sin2|k|z\left\{ 4|k|^{2}\right.\right.\\
 & \times & \left(\Delta k_{b}\left(G_{a-}^{*}-1\right)-\Delta k_{a}\left(G_{a-}^{*}+1\right)+\Delta k_{b}\right)\\
 & - & \left(\Delta k_{b}^{2}\left(\Delta k_{b}-2\Delta k_{a}\right)+\left(\Delta k_{ab}^{3}-2\Delta k_{a}^{2}\Delta k_{b}\right)\right.\\
 & \times & \left.\left.\left(G_{a-}^{*}-1\right)\right)\right\} -i\Delta k_{a}^{2}\Delta k_{b}^{2}\Delta k_{ab}^{3}\left(G_{a+}^{*}-1\right)\\
 & \times & \sin2|k|z+16|k|^{5}\left(-\Delta k_{b}\Delta k_{ab}G_{a-}^{*}\cos2|k|z\right.\\
 & + & \left(\Delta k_{a}\left(G_{ab-}^{*}-1\right)+\Delta k_{ab}\left(G_{a+}^{*}-1\right)+\Delta k_{b}\right)\\
 & \times & \left.\Delta k_{a}\right)-4|k|^{3}\left\{ \Delta k_{b}\Delta k_{ab}\cos2|k|z\right.\\
 & \times & \left(\Delta k_{a}^{2}\left(3G_{a+}^{*}-2\right)-\Delta k_{a}\Delta k_{b}G_{a+}^{*}-\Delta k_{b}^{2}G_{a-}^{*}\right)\\
 & + & \Delta k_{a}\left(\left(\Delta k_{b}^{2}\Delta k_{ab}+\Delta k_{ab}^{3}\right)\left(G_{a+}^{*}-1\right)\right.\\
 & + & \Delta k_{b}^{3}+\Delta k_{b}\Delta k_{ab}^{2}-\left(G_{ab+}^{*}-1\right)\\
 & \times & \left.\left.\left(2\Delta k_{a}^{2}\Delta k_{b}-4\Delta k_{a}\Delta k_{b}^{2}+\Delta k_{a}^{3}+2\Delta k_{b}^{3}\right)\right)\right\} \\
 & - & |k|\Delta k_{a}\Delta k_{b}\Delta k_{ab}^{2}\left(\Delta k_{b}\Delta k_{ab}\left(G_{a-}^{*}-1\right)\right.\\
 & + & 2\Delta k_{a}^{2}\left(G_{ab+}^{*}-1\right)-\Delta k_{b}^{2}+\cos2|k|z\left(-\Delta k_{b}^{2}\right.\\
 & + & \left.\left.\left.\left(\Delta k_{a}\Delta k_{b}-2\Delta k_{a}^{2}+\Delta k_{b}^{2}\right)\left(G_{a+}^{*}-1\right)\right)\right)\right],
\end{array}
\]

\begin{equation}
 \begin{array}{lcl}
l_{11} & = & -\frac{2C_{ab}k\left(G_{ab+}-1\right)}{k^{*}}\left[32|k|^{6}\left(\Delta k_{a}\left(G_{ab-}^{*}-1\right)+\Delta k_{b}\right.\right.\\
 & + & \left.\Delta k_{ab}\left(G_{a+}^{*}-1\right)\right)+16i\Delta k_{b}\Delta k_{ab}|k|^{5}G_{a-}^{*}\sin2|k|z\\
 & - & 2\Delta k_{a}^{2}\Delta k_{b}^{2}\Delta k_{ab}^{3}\left(G_{a+}^{*}-1\right)\sin^{2}|k|z+\left\{ 4i\Delta k_{b}\Delta k_{ab}\right.\\
 & \times & |k|^{3}\left(-\Delta k_{a}\Delta k_{b}G_{a+}^{*}-\Delta k_{b}^{2}G_{a-}^{*}+\Delta k_{a}^{2}\left(3G_{a+}^{*}-2\right)\right)\\
 & - & i\Delta k_{a}\left(\Delta k_{b}^{2}+\left(2\Delta k_{a}^{2}-\Delta k_{a}\Delta k_{b}-\Delta k_{b}^{2}\right)\left(G_{a+}^{*}-1\right)\right)\\
 & \times & \left.\Delta k_{b}\Delta k_{ab}^{2}|k|\right\} \sin2|k|z-8|k|^{4}\left\{ \Delta k_{b}\Delta k_{ab}\cos2|k|z\right.\\
 & \times & \left(\Delta k_{b}G_{a-}^{*}-\Delta k_{a}\left(G_{a-}^{*}+1\right)\right)-\Delta k_{a}\left(G_{ab+}^{*}-1\right)\\
 & \times & \left(\Delta k_{a}\Delta k_{b}+3\Delta k_{ab}^{2}\right)+\Delta k_{b}\left(\Delta k_{ab}^{2}+\Delta k_{b}^{2}\right)+\left(G_{a+}^{*}\right.\\
 & - & \left.\left.1\right)\left(-5\Delta k_{a}^{2}\Delta k_{b}+4\Delta k_{a}\Delta k_{b}^{2}+3\Delta k_{a}^{3}-2\Delta k_{b}^{3}\right)\right\} \\
 & + & 2\Delta k_{ab}|k|^{2}\left\{ \Delta k_{b}\left(\Delta k_{b}^{2}\left(\Delta k_{b}-2\Delta k_{a}\right)+\left(G_{a+}^{*}-1\right)\right.\right.\\
 & \times & \left.\left(\Delta k_{ab}^{3}-2\Delta k_{a}^{2}\Delta k_{b}\right)\right)\cos2|k|z+\Delta k_{a}^{3}\left(G_{ab-}^{*}+1\right)\\
 & \times & \left(2\Delta k_{ab}-\Delta k_{b}\right)+\Delta k_{b}^{3}\Delta k_{ab}+\left(G_{a+}^{*}-1\right)\\
 & \times & \left.\left.\left(2\Delta k_{a}^{2}\Delta k_{ab}^{2}-2\Delta k_{a}\Delta k_{b}^{3}+3\Delta k_{a}^{2}\Delta k_{b}^{2}+\Delta k_{b}^{4}\right)\right\} \right],\\
l_{12} & = & \frac{-2C_{ab}k\left(G_{ab+}-1\right)\left(4|k|^{2}-\Delta k_{ab}^{2}\right)}{k^{*}}\left[8|k|^{4}\left(\Delta k_{a}\left(G_{ab-}^{*}-1\right)\right.\right.\\
 & + & \left.\Delta k_{ab}\left(G_{a+}^{*}-1\right)+\Delta k_{b}\right)+\Delta k_{a}^{2}\Delta k_{b}^{2}\Delta k_{ab}\sin^{2}|k|z\\
 & \times & \left(G_{a+}^{*}-1\right)-2|k|^{2}\left\{ \Delta k_{a}\Delta k_{ab}\left(G_{a+}^{*}-1\right)\cos2|k|z\right.\\
 & \times & \Delta k_{b}+\Delta k_{ab}\left(\Delta k_{a}^{2}+\Delta k_{b}^{2}\right)\left(G_{a+}^{*}-1\right)+\Delta k_{a}^{3}\\
 & \times & \left.\left(G_{a-}^{*}-1\right)\Delta k_{b}^{3}\right\} +i\Delta k_{a}\Delta k_{b}|k|\left(\Delta k_{a}^{2}-\Delta k_{b}^{2}\right)\\
 & \times & \left.\left(G_{a+}^{*}-1\right)\sin2|k|z\right],\\
l_{13} & = & \frac{-2C_{ab}k|k|\left(G_{ab+}-1\right)}{k^{*}}\left[32|k|^{5}\left(\Delta k_{a}\left(G_{ab-}^{*}-1\right)+\Delta k_{b}\right.\right.\\
 & + & \left.\Delta k_{ab}\left(G_{a+}^{*}-1\right)\right)-4i\Delta k_{b}\Delta k_{ab}|k|^{2}\sin2|k|z\\
 & \times & \left\{ 4|k|^{2}G_{a-}^{*}-\Delta k_{a}\Delta k_{b}\left(3G_{a+}^{*}-2\right)+\Delta k_{a}^{2}G_{a+}^{*}\right.\\
 & - & \left.\Delta k_{b}^{2}G_{a-}^{*}\right\} -i\Delta k_{a}\Delta k_{b}^{2}\Delta k_{ab}^{2}\sin2|k|z\\
 & \times & \left(-\Delta k_{b}+\Delta k_{ab}\left(G_{a+}^{*}-1\right)\right)-8|k|^{3}\left\{ \Delta k_{b}\Delta k_{ab}\right.\\
 & \times & \left(-\Delta k_{b}G_{a-}^{*}+\Delta k_{a}\left(G_{a+}^{*}+1\right)\right)\cos2|k|z\\
 & - & \Delta k_{ab}\left(G_{ab+}^{*}-1\right)\left(\Delta k_{a}^{2}+\Delta k_{b}\Delta k_{ab}\right)\\
 & + & \left.\left(\Delta k_{b}+\Delta k_{ab}\left(G_{a+}^{*}-1\right)\right)\left(\Delta k_{b}^{2}+\Delta k_{ab}^{2}\right)\right\} \\
 & - & 2\Delta k_{b}\Delta k_{ab}|k|\left\{ \cos2|k|z\left(-\Delta k_{b}^{2}\left(\Delta k_{a}+\Delta k_{ab}\right)\right.\right.\\
 & - & \left.\Delta k_{ab}^{2}\left(\Delta k_{a}+\Delta k_{b}\right)\left(G_{a+}^{*}-1\right)\right)-\Delta k_{b}\Delta k_{ab}^{2}\\
 & \times & \left.\left.\left(G_{a+}^{*}-1\right)+\Delta k_{a}^{3}\left(G_{ab+}^{*}-1\right)-\Delta k_{b}^{2}\Delta k_{ab}\right\} \right],\\
l_{14} & = & \frac{-2C_{ab}k^{2}\left(G_{ab+}-1\right)}{k^{*}}\left[2\Delta k_{a}\Delta k_{b}^{2}\Delta k_{ab}^{2}\left(-\Delta k_{b}+\Delta k_{ab}\right.\right.\\
 & \times & \left.\left(G_{a+}^{*}-1\right)\right)+2i\Delta k_{b}\Delta k_{ab}|k|\sin2|k|z\left\{ -4|k|^{2}\right.\\
 & \times & \left(-\Delta k_{b}G_{a-}^{*}+\Delta k_{a}\left(G_{a+}^{*}+1\right)\right)+\left(\Delta k_{a}+\Delta k_{ab}\right)\\
 & \times & \left.\Delta k_{b}^{2}+\Delta k_{ab}^{2}\left(\Delta k_{a}+\Delta k_{b}\right)\left(G_{a+}^{*}-1\right)\right\} +16|k|^{4}\\
 & \times & \left\{ -\Delta k_{b}\Delta k_{ab}G_{a-}^{*}\cos2|k|z+\Delta k_{a}\left(\left(G_{a+}^{*}-1\right)\right.\right.\\
 & \times & \left.\Delta k_{ab}+\left(\left(\Delta k_{b}-\Delta k_{ab}\right)\left(G_{ab+}^{*}-1\right)-\Delta k_{b}\right)\right\} \\
 & - & 4|k|^{2}\left\{ \Delta k_{b}\Delta k_{ab}\left(-\Delta k_{a}\Delta k_{b}\left(3G_{a+}^{*}-2\right)-\Delta k_{b}^{2}\right.\right.\\
 & \times & \left.G_{a-}^{*}+\Delta k_{a}^{2}G_{a+}^{*}\right)\cos2|k|z+\Delta k_{a}\left(\left(\Delta k_{b}-\Delta k_{ab}\right)\right.\\
 & \times & \Delta k_{a}^{2}\left(G_{ab+}^{*}-1\right)+\left(\Delta k_{b}^{2}+\Delta k_{ab}^{2}\right)\left(-\Delta k_{b}\right.\\
 & + & \left.\left.\left.\left.\Delta k_{ab}\left(G_{a+}^{*}-1\right)\right)\right)\right\} \right]
\end{array}\label{eq:coefficient-l}
\end{equation}

with $G_{i\pm}=\left(1\pm\exp(-i\Delta k_{i}z)\right)$ for $i\in\left\{ a,b,ab\right\}$,
and $\Delta k_{ab}=\Delta k_{a}-\Delta k_{b}$. Also, $C_{a}=\frac{\Gamma_{a}}{\left[4|k|^{2}-\Delta k_{a}^{2}\right]}$, $C_{b}=\frac{\Gamma_{b}}{\left[4|k|^{2}-\Delta k_{b}^{2}\right]}$ and $C_{ab}=\frac{\Gamma_{a}^{*}\Gamma_{b}}{\Delta k_{a}\Delta k_{b}\Delta k_{ab}\left[4|k|^{2}-\Delta k_{a}^{2}\right]\left[4|k|^{2}-\Delta k_{b}^{2}\right]\left[4|k|^{2}-\Delta k_{ab}^{2}\right]}$.

\textbf{Acknowledgement:} J P thanks Project LO1305 of the Ministry
of Education, Youth and Sports of the Czech Republic\textcolor{blue}{.}

\end{document}